\documentclass[prd,aps,eqsecnum,floatfix,nofootinbib,preprint,tightenlines]{revtex4-2}

\usepackage{latexsym}
\usepackage{graphicx}
\usepackage{multirow}
\usepackage[dvipsnames]{xcolor}
\usepackage{hyperref}

\def\ttl#1{{\it #1}}
\def\floatcaption#1#2{ \caption{#2 \label{#1}} }

\def\bibi{\bibitem}

\let\inodot=\i

\def\a{\alpha}
\def\b{\beta}
\def\c{\chi}
\def\d{\delta}
\def\g{\gamma}

\def\i{\iota}

\def\m{\mu}
\def\n{\nu}

\def\p{\pi}                     % Also, \varpi
\def\r{\rho}                    %       \varrho
\def\s{\sigma}                  %       \varsigma
\def\t{\tau}

\def\D{\Delta}

\def\G{\Gamma}

\def\P{\Pi}

\def\cc{{\cal C}}

\def\co{{\cal O}}

\def\cbo{{\,\raise-.15ex\Sc [\,}}                       % curly "
\def\ddt#1{{\buildrel {\hbox{\LARGE .\kern-2pt.}} \over {#1}}}% double dot-over
\def\ie{\mbox{\it i.e.}}
\def\eg{\mbox{\it e.g.}}
\def\etc{\mbox{\it etc.}}

\def\half{{1\over 2}}

\def\ttl#1{{\it #1}}
\def\seef{{\it cf.\  }}

\def\ansatz{{\it ansatz}}

\begin{document}

\begin{boldmath}
\begin{center}
{\large{\bf
The strong coupling from an improved $\tau$ vector isovector\\ spectral
function}
}\\[8mm]
Diogo Boito,$^a$ Maarten Golterman,$^b$  Kim Maltman,$^{c,d}$ Santiago Peris,$^e$\\ 
Marcus V. Rodrigues$^a$ and Wilder Schaaf$^{\,b}$\\[8 mm]
$^a$Instituto de F{\'\inodot}sica de S{\~a}o Carlos, Universidade de S{\~a}o Paulo\\
CP 369, 13570-970, S{\~a}o Carlos, SP, Brazil
\\[5mm]
$^b$Department of Physics and Astronomy, San Francisco State University,\\
San Francisco, CA 94132, USA\\
[5mm]
$^c$Department of Mathematics and Statistics,
York University\\  Toronto, ON Canada M3J~1P3
\\[5mm]
$^d$CSSM, University of Adelaide, Adelaide, SA~5005 Australia
\\[5mm]
$^e$Department of Physics and IFAE-BIST, Universitat Aut\`onoma de Barcelona\\
E-08193 Bellaterra, Barcelona, Spain
\\[10mm]
\end{center}
\end{boldmath}
\begin{quotation}
We combine ALEPH and OPAL results for the spectral distributions 
measured in $\t\to\p^-\p^0\n_\t$, $\t\to2\p^-\p^+\p^0\n_\t$ and 
$\t\to\p^-3\p^0\n_\t$ decays with (i) recent BaBar results for the 
analogous $\t\to K^- K^0\n_\t$ distribution and (ii) estimates of 
the contributions from other hadronic $\t$-decay modes obtained using 
CVC and electroproduction data, to obtain a new and more precise 
non-strange, inclusive vector, isovector spectral function. 
The BaBar $K^- K^0$ and CVC/electroproduction results provide us 
with alternate, entirely data-based input for the contributions of all 
exclusive modes for which ALEPH and OPAL employed Monte-Carlo-based 
estimates. We use the resulting spectral function to determine 
$\a_s(m_\t)$, the strong coupling at the $\t$ mass scale, employing
finite energy sum rules. Using the fixed-order perturbation theory (FOPT)
prescription, we find $\a_s(m_\t)=0.3077\pm 0.0075$, which 
corresponds to the five-flavor result $\a_s(M_Z)=0.1171\pm 0.0010$ 
at the $Z$ mass. While we also provide an estimate using contour-improved 
perturbation theory (CIPT), we point out that the FOPT prescription is 
to be preferred for comparison with other $\a_s$ determinations employing 
the $\overline{{\rm MS}}$ scheme, especially given the inconsistency 
between CIPT and the standard operator product expansion recently pointed 
out in the literature. Additional experimental input on the dominant 
$2\pi$ and $4\pi$ modes would allow for further improvements to the current 
analysis.

\end{quotation}

\newpage
\section{\label{intro} Introduction}
Since the calculation of the term of order $\a_s^4$ in the Adler 
function \cite{PT}, there has been a revived interest in the
 determination of the strong coupling, $\a_s(m_\t)$, at the
$\t$ mass scale $m_\t$, from non-strange hadronic $\t$ decays. Two LEP 
experiments, ALEPH \cite{ALEPH,ALEPH2,ALEPH13} and OPAL \cite{OPAL} conducted 
measurements of hadronic $\t$ decays from which the inclusive non-strange 
vector ($V$) and axial ($A$) isovector spectral functions were 
extracted with high accuracy as a function of the squared invariant mass 
$s$. A third experiment, CLEO, also used $V$ and $A$ inclusive spectral 
data in an early determination of $\a_s(m_\t)$ \cite{CLEO}, but these 
data have not been made publicly available.\footnote{To the best of our 
knowledge, only data for the decay $\t\to\p^-\p^0\n_\t$ are publicly 
available \cite{CLEO2pi}.}

Most determinations of $\a_s$ since Ref.~\cite{PT} have been based on the 
ALEPH data, the most recent of these using the 2013 version of this 
data \cite{ALEPH13,alphas14,Pich}, in which an earlier 
problem \cite{ustau10} with the data covariance matrix was 
corrected \cite{ALEPH13}. An exception is the determination of $\a_s$ 
in Ref.~\cite{alphas1}, which was based on the OPAL data. This was later 
updated in Ref.~\cite{alphas2} for changes in the exclusive-mode $\tau$ 
branching fractions (BFs) since the original 1998 OPAL publication.
The determinations based on the ALEPH \cite{alphas14} and OPAL \cite{alphas2} 
data are consistent with each other, leading Ref.~\cite{alphas14} to quote a 
weighted average as the best result taking the determination from both 
data sets into account. 
While this weighted average should be reliable, it was not based on a
fit of the combined ALEPH and OPAL data, and compatibility of the
two $\a_s$ values does not test the compatibility of the two data sets 
directly.\footnote{It is, in any case, important to update and complement 
these data with more recent experimental results, where available.}

Clearly, what one would like to do instead is to combine the two data 
sets to produce a single data set whose spectral functions and corresponding
covariance matrices reflect locally, \ie, in an $s$-dependent manner, 
the combined constraints of the ALEPH and OPAL data. The process of
combining the data sets tests for their compatibility, and the result is 
a data set with smaller errors than either of the two data sets alone.
In this sense, the combined spectral functions will be the ``best available'' 
extracted from hadronic $\t$ decays. A number of hadronic quantities
useful for hadron phenomenology, such as $\a_s$, certain low-energies 
constants of chiral perturbation theory and operator product expansion 
condensates can then be determined from the $V$ and $A$ spectral functions. 

In this paper, we begin this program by constructing the combined inclusive
non-strange $V$ spectral function, and using this to obtain the most precise 
value of the strong coupling that can be obtained from the combined 
ALEPH and OPAL hadronic $\t$-decay data in the $V$ channel, supplemented 
with $e^+e^- \to hadrons$ cross-section data for some small but 
non-negligible residual exclusive modes. The process of combining the 
two spectral functions involves several steps, 
and includes updating the normalizations of exclusive channels using 
updated BFs, before the two data sets are combined. 

There are two reasons for limiting ourselves to the $V$ spectral function 
in this paper. The first is that the $V$ channel is dominated by 
the $2\p$ and $4\p$ decay modes, while the remaining channels,
including those for which ALEPH and OPAL used Monte-Carlo (MC) input, 
play a much smaller relative role in the $V$ than in the $A$ channel. 
For OPAL, all $V$-channel modes apart from the dominant $2\pi$ and $4\pi$ 
channels, $\pi^- \pi^0$, $\pi^- 3\pi^0$ and $2\pi^-\pi^+\pi^0$, are listed 
as having a MC source~\cite{OPAL}. For ALEPH, MC simulations are used 
for the $K^- K^0$, $K\overline{K} 2\pi$ and $6\pi$ $V$ contributions with the 
simulation also used for the part of the $\omega\pi^-$ distribution not
reconstructed in the $2\pi^-\pi^+\pi^0$ mode~\cite{ALEPH,ALEPH2,ALEPH13}. 

A second important reason for focusing on the $V$ channel is that the CVC 
(conserved vector current) relation\footnote{The notation ``CVC'' 
reflects the observation that the charged $V$ current responsible for
non-strange hadronic $\t$ decays is the charged member of the same isospin 
multiplet as the $I=1$ part of the electromagetic current.} between 
the $\t$-based $V$ spectral function and $I=1$ hadronic $e^+ e^-$ cross 
section contributions allows almost all of the smaller, but still 
numerically relevant, contributions from exclusive modes other than 
the dominant $2\pi$ and $4\pi$ ones to be significantly improved using 
recent high-precision exclusive-mode $e^+ e^- \rightarrow hadrons$ 
cross-section data. The use of CVC and electroproduction data paves 
the way for an almost fully experimental, and in this sense improved, 
determination of the $V$ spectral function.
Such CVC improvements are, of course, not possible in the $A$ 
channel, which, in addition, receives larger relative contributions from 
higher-multiplicity exclusive modes for which exclusive-mode spectral 
function contributions and covariances were not provided by OPAL and ALEPH. 

For these reasons, we postpone a discussion of the $A$ case to a future work. 
As far as the determination of $\alpha_s$ is concerned, we found, in 
previous work, that, while the addition of the $A$ channel to the analysis 
provided a nice consistency check~\cite{alphas14,alphas1,alphas2}, it did 
not help reduce the error on $\a_s$. We thus consider the determination of 
$\a_s$ using an updated version of the $V$ spectral function alone to be 
of interest.

There are thus two parts to the work reported in this paper. In the first 
part, we update the determination of the inclusive non-strange $V$ spectral 
function. This itself involves two steps. First, we combine, 
and hence update, the results for the contributions from the exclusive 
modes for which ALEPH and OPAL data is publicly available (the dominant 
$2\pi$ and $4\pi$ modes), using the method employed to combine 
exclusive-mode $e^+ e^- \rightarrow hadrons$ $R(s)$ data from different 
experiments and described in detail in Ref.~\cite{KNT18}. Second, we use 
recent results for $\tau\rightarrow K^- K^0\nu_\tau$, together 
with CVC and recent exclusive-mode $e^+ e^-\rightarrow hadrons$ cross 
section results, to improve the determination of the contributions from 
the remaining modes, which in this paper we will refer to as the 
``residual'' modes. In the second part of the paper, we apply the 
strategy developed in Refs.~\cite{alphas14,alphas1,alphas2} to extract 
$\a_s(m_\t)$ from the improved inclusive $V$ non-strange spectral 
function obtained in the first part.

The determination of $\a_s$ from $V$ and/or $A$ spectral functions makes 
use of finite energy sum rules (FESRs), which allow us to relate $\a_s$ at 
the $\t$ scale through the Adler function, calculated in QCD perturbation 
theory, to integrals of the spectral function from threshold to the 
$\t$ mass \cite{shankar,MPR,CK,CKT,KPT,FNR,BLR,Braaten88,BNP}. Even so, 
with the strong coupling at scales around the $\t$ mass being rather 
large, these sum rules are ``contaminated'' by non-perturbative effects. 
These non-perturbative effects are clearly visible in the experimental 
data, since the shape of the experimental inclusive spectral function 
does not agree with perturbation theory. As a result of the presence
of resonances, the experimental spectral functions oscillate around 
the perturbative predictions, with these oscillations remaining visible for
$s$ close to $m_\t^2$. Several methods have been designed for dealing with 
these non-perturbative effects. The oldest 
method (the ``truncated OPE'' strategy) employs weight functions assumed 
to suppress the effects of the observed ``duality violating'' resonant
oscillations and assumes the reliability of a truncation in dimension of 
operator product expansion (OPE) contributions required to make the 
analysis practical \cite{BNP,DibPich}. A newer method, the ``DV-model 
strategy,'' instead takes quark-hadron duality violations (DVs), \ie,
collective resonance effects, into account, and requires only very 
mild assumptions about the behavior of the OPE \cite{alphas1,BCGMP}. 
While it is not straightforward to ascertain the reliability of estimates 
of non-perturbative effects, self-consistency tests show that the truncated 
OPE strategy leads to unreliable results, with a theoretical uncertainty 
arising from the neglect of higher-order OPE terms and of DV contributions 
that is not accounted for~\cite{critical,EManalysis,conf}. 
We will thus employ the DV-model strategy, which was first developed in 
Ref.~\cite{alphas1} and thoroughly tested in Refs.~\cite{alphas14,alphas1,alphas2}, 
and for which to date no inconsistencies have been found.\footnote{Criticism
of the DV-model strategy in Ref.~\cite{Pich} was refuted in Ref.~\cite{critical}.}

This paper is organized as follows. In Sec.~\ref{theory} we give a brief 
overview of FESRs, \ie, the theory needed to extract $\a_s$ from 
spectral function input. Then, in Sec.~\ref{data}, we review the ALEPH 
and OPAL data sets, describe in detail how we combine their publicly 
available results for the dominant $2\pi$ and $4\pi$ modes, and outline 
the use of new $\tau\rightarrow K^- K^0\nu_\tau$ data and CVC to improve 
the determination of contributions from the remaining residual $V$ exclusive 
modes. In Sec.~\ref{strong coupling} we present the results 
for $\alpha_s(m_\tau )$ obtained from DV-model-strategy-based fits to 
our updated, inclusive, non-strange $I=1$, $V$ spectral function. Finally, 
Sec.~\ref{conclusion} contains our conclusions and prospects for future 
progress.

\section{\label{theory} Theory overview}
In Sec.~\ref{FESR}, we briefly recapitulate the use of FESRs to extract 
$\a_s$ from spectral function input and define our theoretical framework. The 
choice of sum rules employed in our fits is discussed in Sec.~\ref{weightOPE}. 
In Sec.~\ref{FOPTCIPT} we argue that fixed-order perturbation theory (FOPT)
should be favored over contour-improved perturbation theory (CIPT) 
\cite{CIPT,CIPT2}, if the goal is to convert our result for $\a_s(m_\t)$ 
to a value at the $Z$ mass to be compared to $\overline{\rm MS}$ values 
of $\a_s(M_Z)$ obtained from other sources. We also explain how we 
estimate the systematic error associated with the necessary truncation 
of perturbation theory beyond order $\a_s^4$.

\subsection{\label{FESR} Finite energy sum rules}
The sum-rule analysis starts from the current-current correlation functions
\begin{eqnarray}
\label{correl}
\P_{\m\n}(q)&=&i\int d^4x\,e^{iqx}\langle 0|T\left\{J_\m(x)J^\dagger_\n(0)\right\}|0\rangle\\
&=&\left(q_\m q_\n-q^2 g_{\m\n}\right)\P^{(1)}(q^2)+q_\m q_\n\P^{(0)}(q^2)\nonumber\\
&=&\left(q_\m q_\n-q^2 g_{\m\n}\right)\P^{(1+0)}(q^2)+q^2 g_{\m\n}\P^{(0)}(q^2)\ ,\nonumber
\end{eqnarray}
where $J_\m$ stands for the non-strange vector ($V$) current $\overline{u}\g_\m d$
or axial ($A$) current $\overline{u}\g_\m\g_5 d$, and
the superscripts $(0)$ and $(1)$ label spin. The decomposition in the
third line is useful because $\P^{(1+0)}(q^2)$ and $q^2\P^{(0)}(q^2)$
are free of kinematic singularities.  With $s=q^2$, the spectral function
\begin{equation}
\label{spectral}
\r^{(1+0)}(s)=\frac{1}{\p}\;\mbox{Im}\,\P^{(1+0)}(s)\ ,
\end{equation}
and the known analytical properties of $\P^{(1+0)}(z)$,
application of Cauchy's theorem to the contour in Fig.~\ref{cauchy-fig} implies the FESR
\begin{equation}
\label{cauchy}
I^{(w)}_{V/A}(s_0)\equiv\frac{1}{s_0}\int_0^{s_0}ds\,w(s)\,\r^{(1+0)}_{V/A}(s)
=-\frac{1}{2\p i\, s_0}\oint_{|z|=s_0}
dz\,w(z)\,\P^{(1+0)}_{V/A}(z)\ .
\end{equation}
The sum rule is valid for any $s_0>0$ and any weight $w(s)$ analytic 
inside and on the contour \cite{shankar,MPR,CK,CKT,KPT,FNR,BLR}. In this 
paper, we will always choose $w(z)$ to be polynomial in $z$.

%%%%%%%%%%%%%%%%%%%
\begin{figure}
\vspace*{4ex}
\begin{center}
\includegraphics*[width=6cm]{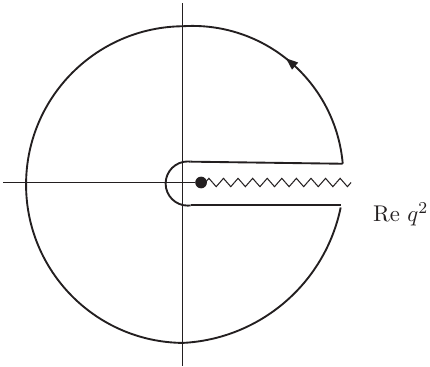}
\end{center}
\begin{quotation}
\floatcaption{cauchy-fig}%
{{\it Analytic structure of $\P^{(1+0)}(q^2)$ in the complex $s=q^2$ plane.
There is a cut on the positive real axis starting at $s=q^2=4m_\p^2$ (a pole 
at $s=q^2=m_\p^2$ and a cut starting at $s=9m_\p^2$) for the $V$ ($A$)
case. The solid curve shows the contour used in Eq.~(\ref{cauchy}).}}
\end{quotation}
\vspace*{-4ex}
\end{figure}
%%%%%%%%%%%%%%%%%%%

The flavor-$ud$ $V$ and $A$ spectral functions can be
experimentally determined from the differential versions of the ratios
\begin{equation}
\label{R}
R_{V/A;ud}=
{\frac{\G [\t\rightarrow ({\rm hadrons})_{V/A;ud}\n_\t (\g ) ]}
{\G [\t\rightarrow e\bar{\n}_e \nu_\tau (\g ) ]}}\ ,
\end{equation}
of the hadronic decay width induced by the relevant current to that
for the electron mode. Explicitly~\cite{tsai71},
\begin{equation}
\label{taukinspectral}
{\frac{dR_{V/A;ud}(s)}{ds}}= 12\pi^2\vert V_{ud}\vert^2 S_{EW}\,
{\frac{1}{m_\tau^2}} \left[ w_T(s;m_\tau^2) \rho_{V/A;ud}^{(1+0)}(s)
- w_L(s;m_\tau^2) \rho_{V/A;ud}^{(0)}(s) \right]\ ,
\end{equation}
where $S_{EW}$ is a short-distance electroweak correction and
\begin{eqnarray}
\label{kinweights}
w_T(s;s_0)&=&\left(1-\frac{s}{s_0}\right)^2\left(1+2\,\frac{s}{s_0}\right)\ ,\\
w_L(s;s_0)&=&2\,\frac{s}{s_0}\left(1-\frac{s}{s_0}\right)^2\ .\nonumber
\end{eqnarray}
With the exception of the known pion-pole part of $\rho_{A;ud}^{(0)}$,
the spectral functions $\rho_{V/A;ud}^{(0)}$ are chirally suppressed,
$\rho_{V/A;ud}^{(0)}(s) = O[(m_d\mp m_u)^2]$, and can be neglected in 
the non-strange case. The spectral functions
$\rho^{(1+0)}_{V/A;ud}(s)$ can thus be determined directly
from $dR_{V/A;ud}(s)/ds$ for any positive value of $s\le m_\t^2$, allowing 
us to apply the FESR~(\ref{cauchy}) for arbitrary $s_0\le m_\t^2$ and 
arbitrary analytic weight $w(s)$ to the data.
As in Ref.~\cite{alphas14}, we will denote the experimental
spectral integral on the left-hand side of Eq.~(\ref{cauchy}) by
$I_{V/A;\rm ex}^{(w)}(s_0)$ and the theoretical
representation of the contour integral on the right-hand side by
$I_{V/A;\rm th}^{(w)}(s_0)$.

For large enough $|s|=s_0$, and sufficiently far away from the positive 
real axis, $\P^{(1+0)}(s)$ can be approximated by the OPE
\begin{equation}
\label{OPE}
\P^{(1+0)}_{\rm OPE}(z)=\sum_{k=0}^\infty \frac{C_{2k}(z)}{(-z)^{k}}\ ,
\end{equation}
where the logarithmic $z$ dependence of the OPE coefficients $C_{2k}$ can be
calculated in perturbation theory.

For the $k=0$ term, it is convenient to consider, instead of $\P(z)$, the 
Adler function $D(z) \equiv -z\,d\P(z)/dz$, which is finite and independent 
of the renormalization scale $\m$. Accordingly, the $k=0$ contribution to 
the right-hand side of Eq.~(\ref{cauchy}) can be expressed in terms of the
Adler function via partial integration.
The $k=0$ contribution $D_0(z)$ to $D(z)$ takes the form
\begin{equation}
\label{pertth}
D_0(z)\equiv -z\,\frac{dC_0(z)}{dz}=\frac{1}{4\p^2}\sum_{n=0}^\infty\left(\frac{\a_s(\m^2)}{\p}\right)^n\sum_{m=1}^{n+1}
mc_{nm}\left(\log\frac{-z}{\m^2}\right)^{m-1}\ ,
\end{equation}
where the coefficients $c_{nm}$ are known to order $\a_s^4$~\cite{PT}.
The independence of $D_0(z)$ on $\m$ implies that only the coefficients 
$c_{n1}$ are independent; the $c_{nm}$ with $m>1$ can be expressed in 
terms of $c_{n1}$ through use of the renormalization group.\footnote{See 
for instance Ref.~\cite{MJ}.} In the $\overline{\rm MS}$ scheme, 
$c_{01}=c_{11}=1$, $c_{21}=1.63982$, $c_{31}=6.37101$
and $c_{41}=49.07570$, for three flavors \cite{PT}. While $c_{51}$ is 
not known at present, we will use the estimate $c_{51}=283$ of Ref.~\cite{BJ}, 
with a 50\% uncertainty. For the running of $\a_s$ we use the four-loop 
$\overline{\rm MS}$ $\b$-function, but we have checked that using five-loop 
running instead \cite{5loop,5loop2} leads to differences of order $10^{-4}$ 
or less in our results for $\a_s(m_\t)$.

The $C_{2k}$ with
$k\ge 1$ are different for the $V$ and $A$ channels, and, for $k>1$,
contain non-perturbative $D=2k$ condensate contributions. As in
Refs.~\cite{alphas1,alphas2}, we will neglect purely perturbative quark-mass
contributions to $C_{2k}$, $k\ge 1$, as they are numerically very 
small for the non-strange FERSs we consider in this paper. We
will also neglect the $z$-dependence of the coefficients $C_{2k}$ for $k>1$. 
For a more detailed discussion of our treatment of the $D>0$ OPE
contributions, we refer to Ref.~\cite{alphas1}.

Perturbation theory, and in general the OPE, breaks down near the positive 
real $s=q^2$ axis \cite{PQW}.  If this were not the case, Eq.~(\ref{cauchy}) 
would establish a direct correspondence between the OPE and the resonant 
behavior of the experimental spectral function, generally referred to as 
quark-hadron duality. We account for the breakdown of this duality by 
replacing the right-hand side of Eq.~(\ref{cauchy}) by
\begin{equation}
\label{split}
-\frac{1}{2\p is_0}\oint_{|z|=s_0}dz\,w(z)\,
\left(\P^{(1+0)}_{\rm OPE}(z)+\D(z)\right)\ ,
\end{equation}
with
\begin{equation}
\label{DVdef}
\D(z)\equiv\P^{(1+0)}(z)-\P^{(1+0)}_{\rm OPE}(z)\ ,
\end{equation}
where $\D(z)\equiv\D_{V/A}(z)$ defines the quark-hadron duality violating 
contribution to $\Pi^{(1+0)}(z)\equiv\Pi_{V/A}^{(1+0)}(z)$.
If $\D(z)$ decays fast enough for $|z|\to\infty$,
Eq.~(\ref{split}) can be rewritten as \cite{CGP}
\begin{equation}
\label{sumrule}
I_{\rm th}^{(w)}(s_0) = -\frac{1}{2\p is_0}\oint_{|s|=s_0}
dz\,w(z)\,\P^{(1+0)}_{\rm OPE}(z)-\frac{1}{s_0}\,
\int_{s_0}^\infty ds\,w(s)\,\frac{1}{\p}\,\mbox{Im}\,
\D(s)\ .
\end{equation}
The imaginary parts $\frac{1}{\p}\,\mbox{Im}\,\D_{V/A}(s)$ can be 
interpreted as the duality-violating parts $\rho_{V/A}^{\rm DV}(s)$ 
of the $V/A$ spectral functions, and represent the resonance-induced, 
oscillatory parts of the spectral functions not captured by the OPE.

In Ref.~\cite{BCGMP}, we developed a theoretical framework for quark-hadron 
duality violations in terms of a generalized Borel--Laplace transform of 
$\P(q^2)$ and hyperasymptotics, building on earlier 
work \cite{russians,russians2,russians3,catalans}. In the chiral limit, 
and assuming that for high energies the spectrum becomes Regge-like in 
the $N_c\to\infty$ limit, we showed that the asymptotic form of 
$\r_{V/A}^{\rm DV}(s)$ for large $s$ can be parametrized as
\begin{equation}
\label{ansatz}
\r_{V/A}^{\rm DV}(s)=\frac{1}{\p}\,\mbox{Im}\,
\D_{V/A}(s)=e^{-\d_{V/A}-\g_{V/A} s}\sin(\a_{V/A}+\b_{V/A} s)\ ,
\end{equation}
up to slowly varying logarithmic corrections in the argument of the 
sine factor, and with $\g\sim 1/N_c$ small but non-zero.\footnote{This 
form was first introduced in Ref.~\cite{CGP05}, and subsequently used in 
Refs.~\cite{alphas14,alphas1,alphas2,CGP,CGPmodel}.} The parameters 
$\b_{V/A}$ are directly related to the Regge slopes in the $V$ and $A$ 
channels, and the parameters $\g_{V/A}$ to the (asymptotic) ratio of the 
width and the mass of the resonances in those channels. While the framework of
Ref.~\cite{BCGMP} is rather general, and the derivation of Eq.~(\ref{ansatz}) 
is based on generally accepted conjectures about QCD (primarily Regge 
behavior), it does not provide a derivation from first principles.
This introduces a certain model dependence in our analysis which, however, 
can be tested by fits to the data. Such tests, in particular, will provide 
information about the values of $s$ above which this asymptotic
form is likely to be sufficiently accurate. We emphasize, however, 
that modifications to the parametrization of Eq.~(\ref{ansatz}) are constrained 
by the general framework of Ref.~\cite{BCGMP}.

Equation~(\ref{ansatz}) introduces, in addition to $\a_s$ and the $D\ge 4$ 
OPE condensates, four new parameters in each channel: 
$\delta_{V/A}, \gamma_{V/A}, \alpha_{V/A}$ and $\beta_{V/A}$. This can 
be compared to the truncated-OPE approach in which DVs are neglected.
Since resonance-induced oscillations are clearly visible in the 
experimental spectral data, and their dynamical effect is comparable in
size to the $\alpha_s$-dependent dynamical effect of perturbative 
corrections to the ($\alpha_s$-independent) parton model 
contribution \cite{critical}, this approach also assumes a model: 
one in which the parameters $\delta_{V/A}$ are effectively set 
to infinity by hand. This is a stronger assumption, and one that 
has been shown to fail a number of subsequent, more stringent 
data-based tests \cite{critical,EManalysis}.

In summary, as in Refs.~\cite{alphas1,alphas2}, we will assume that 
Eq.~(\ref{ansatz}) holds for $s\ge s_{\rm min}$, with $s_{\rm min}$ to be
determined from fits to the data. This assumes of course that the 
$s_{\rm min}$ for which this assumption is valid includes a region
below $m_\t^2$, \ie, that both the OPE~(\ref{OPE}) and the DV 
parametrization~(\ref{ansatz}) can be used in some interval below $m_\t^2$.

\subsection{\label{weightOPE} Choice of weight functions and the OPE}
The logarithmic $s$ dependence of the OPE coefficients $C_D(s)$ is 
calculable in perturbation theory. Because the running of $\a_s$ becomes 
visible only at non-leading order in $\a_s$, this $s$ dependence is an 
$\co(\a_s^2)$ effect in the chiral limit. Such effects were found to be 
safely negligible for $D>0$ in the sum-rule analysis of the OPAL data 
reported in Ref.~\cite{alphas1}, and we will thus ignore them for $D>0$ in 
the present analysis as well. With this simplification, a term in a weight 
$w(z)$ proportional to the monomial $z^n$ picks out the $D>0$ OPE 
contribution with $D=2k=2(n+1)$ in the sum rule~(\ref{cauchy}).\footnote{The 
$D=0$ term, perturbation theory, contributes for all $n$.} The choice of 
a polynomial weight $w(z)$ thus projects the sum rule on a finite number 
of $D>0$ terms in the OPE.

In this paper, we will consider the weights $w(z)=w_n(z/s_0)$ with
\begin{eqnarray}
\label{weights}
w_0(y)&=&1\ ,\\
w_2(y)&=&1-y^2\ ,\nonumber\\
w_3(y)&=&(1-y)^2(1+2y)\ ,\nonumber\\
w_4(y)&=&(1-y^2)^2\ ,\nonumber
\end{eqnarray}
where the subscript indicates the degree of the polynomial. These weights 
explore OPE terms with $D\le 10$, and form a linearly independent basis 
for polynomials up to degree four without a linear term. The weight 
$w_0(y)$ projects only the $D=0$ term of the OPE (\ie, pure perturbation 
theory), the weight $w_2(y)$ projects, in addition, $D=6$, $w_3(y)$ 
projects $D=0$, $D=6$ and $D=8$, while $w_4(y)$ projects $D=0$, $D=6$ and
$D=10$. As the OPE itself diverges as an expansion in $1/z$, it is safer 
to include sum rules with low-degree weights such as $w_0(y)$ and $w_2(y)$ 
in the analysis, and check for consistency among sum rules with different 
weights. We note that $w_3(s/s_0)=w_T(s;s_0)$, \seef\ Eq.~(\ref{kinweights}).

None of these weights contain a term linear in $z$, and thus the 
$D=4$ OPE term does not contribute to the sum rules with these weights.
This choice is motivated by the results of Ref.~\cite{BBJ12}, in which a 
renormalon-model-based study suggested that perturbation theory is
particularly unstable for sum rules with weights containing such a 
linear term.\footnote{Earlier considerations along the same lines can 
be found in Refs.~\cite{alphas1,MJ,BJ}.  The results of Ref.~\cite{BBJ12} 
have been recently corroborated within an alternate approach to 
estimating higher order effects in Ref.~\cite{BO20}.}

The weights $w_{2,3,4}(y)$ are ``pinched,'' \ie, they have zeroes at 
$z=s_0$, and thus suppress contributions from the region near the timelike 
point $z=s_0$ on the contour, and hence also the relative importance of 
integrated DV contributions \cite{KM98,DS99}. The weight $w_2(y)$ has a 
single zero at $z=s_0$ (a single pinch), while the weights $w_3(y)$ 
and $w_4(y)$ are doubly pinched, \ie, have a double zero at $z=s_0$.

\subsection{\label{FOPTCIPT} Perturbative uncertainties and FOPT \textit{\textbf{vs.}} CIPT}
It has become common practice to consider different resummations of 
perturbation theory in order to obtain insight into the effect of 
neglecting terms beyond those explicitly included in evaluating the 
$D=0$ (\ie, perturbative) contribution to the right-hand side of
Eq.~(\ref{cauchy}). The two most commonly used resummation prescriptions 
are fixed-order perturbation theory (FOPT), in which  the scale $\m$ 
in Eq.~(\ref{pertth}) is chosen to be fixed at $\m^2=s_0$, and
contour-improved perturbation theory (CIPT) \cite{CIPT,CIPT2}, a partial 
resummation obtained by choosing $\m^2$ equal to $-z$ point by point 
along the contour. In the CIPT prescription, the coupling is run along the 
contour $|z|=s_0$, using the four- (or five-)loop beta function; as 
a result, only terms with $m=1$ survive in Eq.~(\ref{pertth}). The two 
prescriptions lead to significantly different values of $\a_s$, with 
the difference being comparable to the combination of all other errors 
\cite{alphas14,alphas2}, but more significant, since the results obtained 
with different prescriptions using the same data are highly correlated.

The choice of such a prescription is entangled with the choice of 
renormalization scheme, since the choice of scheme also affects higher 
orders in perturbation theory. Both FOPT and CIPT prescriptions are 
usually considered to be $\overline{\rm MS}$ schemes, but clearly, if 
the choice between FOPT and CIPT is rephrased as a choice of scheme, 
these two schemes are different. Since $\a_s$ is not a physically 
measurable quantity, ideally, one would like to choose a scheme, and 
a prescription, which corresponds to the scheme chosen to quote other 
determinations of $\a_s$ (such as that from $Z$ decay itself), so that 
a direct comparison is possible at $M_Z$.

In our case, the experimental quantities from which we determine $\a_s$ 
are the spectral integrals $I^{(w)}_{\rm ex}(s_0)$, in which $s_0$ is 
varied between $s_0=s_{\rm min}$ and $s_0=m_\t^2$.  Apart from DVs, 
which appear in a transseries beyond the OPE, these quantities are 
then expressed in terms of the OPE, which at $D=0$ is parametrized 
by $\a_s(\m)$ and the ratio of scales $s_0/\m^2$, while at $D\ge 4$ 
also the condensates enter. This leads to the theoretical representations 
$I^{(w)}_{\rm th}(s_0, s_0/\mu^2,\a_s(\m))$, which are then to be equated 
with $I^{(w)}_{\rm ex}(s_0)$. Since $I^{(w)}_{\rm ex}(s_0)$ is an
observable with a single physical scale, $s_0$, the natural choice
is to choose the scale $\m$ equal to the scale of the observable, \ie, to 
choose $\m^2=s_0$.\footnote{While the technical trick that leads to 
the expressions for $I^{(w)}_{\rm th}(s_0)$ involves the contour integral 
over the circle $|z|=s_0$, this trick has little to do with the experimental 
quantities $I^{(w)}_{\rm ex}(s_0)$ from which we determine $\a_s$, and the 
contour can, of course, be deformed to radii larger and/or smaller than 
$s_0$, apart from the endpoints just above and below the timelike axis.}
This corresponds directly to the way the scale is chosen for other 
determinations of $\a_s$. For instance, the hadronic $Z$-decay rate can be 
expressed perturbatively in terms of $\a_s(\m)$ in the $\overline{\rm MS}$ 
scheme, and in that case, the natural choice of scale is $\m=M_Z$. This 
leads us to conclude that, in the case of hadronic $\t$ decays, the 
prescription most directly comparable with other determinations is FOPT.
We emphasize that we do not claim to know which prescription, at a given 
order, gives the best approximation to the QCD answer for 
$I^{(w)}_{\rm ex}(s_0)$. This may depend on the weight $w$, and on the 
order in perturbation theory \cite{BBJ12}. The point is to choose a scheme 
that corresponds most closely to the scheme employed in other
determinations of $\a_s$ at the $Z$-mass. Of course, it is necessary to 
estimate the systematic uncertainty inherent in the truncation of 
perturbation theory, and we will return to this point at the end of 
this subsection.

Before we do this, we present an additional reason for using FOPT as the 
prescription to be used in the FESR determination of $\a_s$ from hadronic 
$\t$ decays. It is well known that perturbation theory for the Adler 
function, and for the quantities $I^{(w)}_{\rm th}(s_0)$, is not convergent.
Attempts to go beyond perturbation theory using Borel resummation
techniques lead to ambiguities in the Borel sum that necessitate the 
introduction of $D>0$ terms in the OPE \cite{MB}. The most well-known 
example is that of the leading infrared renormalon in massless QCD.
This leads to the $D=2k=4$ term in Eq.~(\ref{OPE}), which removes the 
ambiguity associated with this renormalon. The appearance of $D>0$ 
terms in Eq.~(\ref{OPE}) is thus intricately connected to resummations of 
perturbation theory.

In Ref.~\cite{HR}, it was shown that, in general, the FOPT and CIPT series 
lead to different Borel sums, with different analytical properties. 
Moreover, it was pointed out that this different analytical behavior 
of the Borel sums for $I^{(w)}_{\rm th}(s_0)$ appears to invalidate the 
correspondence between infrared renormalons and the $D>0$ terms in the 
OPE in the case of CIPT. 
The Borel sum for the CIPT series does not allow for the
usual renormalon ambiguities that in the case of FOPT
are in one-to-one correspondence with terms in the
OPE.   Thus, there is a mismatch between the use of
CIPT, and the representation of $\Pi^{(1+0)}(z)$
by the OPE, Eq.~(\ref{OPE}).
 This mismatch between the Borel sum and the OPE 
does not happen in FOPT.

This observation casts strong doubts on the consistency of using the 
OPE~(\ref{OPE}) in the case of CIPT. Again, this does not settle the 
issue of whether the (Borel sum of) FOPT or CIPT provides a better 
approximation to QCD. But it does imply that it is theoretically
inconsistent to apply the OPE in the form~(\ref{OPE}) if one uses CIPT 
in the evaluation of the $D=0$ perturbative contributions, and
therefore casts doubts on all CIPT-based extractions of $\a_s$.
We are thus led to the conclusion that FOPT should be taken as the 
preferred choice for analyzing hadronic $\t$ decays using the
OPE. While we will provide determinations of $\a_s(m_\t)$ 
employing both FOPT and CIPT (ignoring the OPE subtleties in the 
case of CIPT) we will quote the FOPT value as our final result, to 
be compared with other $\overline{\rm MS}$ determinations of $\a_s$.
We will also give a CIPT value based on fits to $I^{(w)}_{\rm ex}(s_0)$,
allowing the reader to compare to earlier CIPT results and assess 
the impact of changes in the input inclusive $V$ spectral function.

We will determine the systematic error associated with the use and 
truncation of perturbation theory for our FOPT value of $\a_s(m_\t)$ 
as follows. First, as already indicated in Sec.~\ref{FESR}, we will vary 
the estimated value for $c_{51}=283$ by plus or minus 50\%. This interval 
for $c_{51}$ includes all estimates available in the literature by a rather
wide margin. The variation by $\pm50\%$ was proposed in Ref.~\cite{BJ}, and 
the estimate of Ref.~\cite{PT} falls well inside this interval. Subsequent 
estimates of both the central value and the uncertainty of $c_{51}$ are 
also generously covered by the $\pm 50\%$ variation \cite{BMO,IC}.
This choice of range, of course, provides an estimate only for the 
impact of the uncertainty in the value of the unknown coefficient $c_{51}$, 
and might constitute an underestimate of the total perturbative error.
At present, the perturbative expansion for the Adler function has been 
calculated to order $\a_s^4$. An alternate, potentially more conservative, 
estimate of the perturbative error can thus also be obtained by omitting 
the $\co(\a_s^4)$ contribution altogether, \ie, by setting $c_{4m}=0$ (as 
well as $c_{nm}=0$ for all $n>4$) in Eq.~(\ref{pertth}). Finally, since our 
FOPT determination employs a range of $s_0$ varying between $s_{\rm min}$ 
and $m_\t^2$, a third sensible estimate of the perturbative error can be 
obtained by considering, instead of just $\m^2=s_0$, also the alternate 
choices $\m^2=s_{\rm min}$, $\m^2=m_\t^2$ and $\m^2=2s_0$ for the scale 
$\mu$. We will take the largest of the variations in $\a_s$ obtained by 
applying all three methods above as our best estimate for the systematic 
error associated with the necessary truncation of perturbation theory.
We do not combine the errors obtained by using these three methods, as 
this would correspond to a double-counting of the estimated perturbative 
uncertainties. We will, however, add an independent measure of the 
uncertainty associated with the use of perturbation theory based on the 
comparison of the central values obtained from fits with different 
weights, as explained in more detail in Sec.~\ref{analysis}.

\section{\label{data} Data}
In this section, we construct an updated version of the inclusive, 
non-strange $V$ spectral function, $\rho_{ud;V}(s)$, using publicly available 
ALEPH~\cite{ALEPH,ALEPH2,ALEPH13,ALEPH08} and OPAL~\cite{OPAL} $\tau$ data for 
the contributions of the dominant $2\pi$ and $4\pi$ exclusive modes, recent 
BaBar $\t$-decay results~\cite{babarkkbartau18} for the contribution of the 
$K^-K^0$ mode, and $e^+ e^- \rightarrow hadrons$ cross section data as input
to CVC evaluations of the contributions of the remaining exclusive modes.

In Refs.~\cite{ALEPH,ALEPH2,ALEPH13,OPAL}, the inclusive $V$ and $A$ 
invariant-mass-squared distributions were constructed as sums of
(i) the measured (and publicly available) distributions for the main 
exclusive modes in the channel and (ii) the sum of small contributions 
from the remaining ``residual'' exclusive modes. The publicly available 
exclusive-mode distributions are normalized to then-current 
exclusive-mode BFs. While the accompanying inclusive-sum correlation 
matrices include the contributions from then-current exclusive-mode 
BF uncertainties and correlations for the main exclusive modes, 
the exclusive-mode correlation matrices are provided with the
BF-uncertainty-induced contributions omitted, allowing subsequently 
improved BF information for these modes to be incorporated at 
a later time. 

The $V$ channel modes for which ALEPH and OPAL provide exclusive-mode 
distribution and correlation information are $\p^-\p^0$, $\p^-3\p^0$ and 
$\p^-\p^+\p^-\p^0$. For the $A$ channel, ALEPH provides distribution
and correlation information for only the two $3\pi$ modes, $2\pi^- \pi^+$
and $\pi^- 2\pi^0$, while OPAL provides this information, in addition,
for one of the three $5\pi$ modes, $2\pi^- \pi^+ 2\pi^0$. While all
exclusive-mode distributions are normalized to then-current values of the 
corresponding exclusive-mode BFs, the $s$-dependences of some of the 
residual-mode distributions are taken from Monte Carlo. For OPAL, this 
is true for all but the 
$\pi^- \pi^0 \omega (\rightarrow {\rm \mbox{non-}3}\pi )$ 
residual $A$ mode. For ALEPH, the use of the Monte Carlo simulations is 
explicitly identified as entering the $6\pi$, $K^- K^0$, $K\bar{K}\pi\pi$ 
and $\pi^- \omega (\rightarrow {\rm \mbox{non-}3}\pi )$ contributions to 
the $V$ channel and the $K\bar{K}\pi$ and $K\bar{K}\pi\pi$ contributions to
the $A$ channel. Since the information made publicly available by 
ALEPH and OPAL does not include the individual BF-normalized residual 
exclusive-mode distributions used by the collaborations in determining 
their final inclusive-sum results, it is not possible to update those 
residual-mode contributions to reflect subsequent improvements in our 
knowledge of the exclusive-mode BFs and/or new information on the 
$s$-dependence of exclusive-mode distributions for which Monte Carlo 
was previously employed. Improvements to the residual-mode
contributions must, therefore, come from other sources. The dominant
$V$ and $A$ channel modes (those for which both ALEPH and OPAL 
exclusive-mode distributions are available) represent 98.0\% of 
the inclusive $V$ channel BF and 94.2\% of the continuum inclusive 
$A$ channel BF.

Improving the treatment of residual-mode contributions is much 
easier for the $V$ channel than for the $A$ channel. The reason is
that, strongly motivated by the drive to improve the determination 
of the Standard-Model hadronic vacuum polarization contribution to 
the anomalous magnetic moment of the muon, there has been an intensive 
program of collider and $B$-factory experiments aimed at determining 
the $e^+ e^- \rightarrow hadrons$ cross-sections for all exclusive 
modes contributing to the inclusive $R(s)$ ratio in the region 
below $s\simeq 4$ GeV$^2$. A sizeable fraction of these exclusive
modes can be uniquely classified as either $I=0$ or $I=1$ using 
$G$-parity. The CVC relation between the bare cross section for
the electroproduction of the neutral member, $X^0$, of the exclusive-mode 
isotriplet $X$, $\sigma^b_X(s) \equiv \sigma^b[e^+ e^- \rightarrow X^0]$,
and the contribution of the charged isospin partner, $X^-$, to 
$\rho_{ud;V}(s)$,
\begin{eqnarray}
\left[ \rho_{ud;V}(s)\right]_{X^-}&=&
{\frac{s\, \sigma^b_X(s)}{8\pi^3\alpha_{EM}^2}}
\label{cvc}\end{eqnarray}
then allows the cross section results for those exclusive residual
modes with $I=1$ to be used to determine the corresponding exclusive
residual-mode contributions to $\rho_{ud;V}(s)$. Eq.~(\ref{cvc})
is valid up to isospin-breaking (IB) corrections which, in the absence
of narrow interfering resonances, should be of the order a percent or 
so, and hence numerically negligible on the scale of the experimental
errors on the already small residual-mode contributions. This 
strategy, of using CVC to improve the determination of an otherwise 
poorly determined $V$ exclusive-mode $\tau$ spectral function 
contribution, was pioneered by ALEPH~\cite{ALEPH08}, which used 
the BaBar Dalitz-plot-analysis separation of $I=0$ and $I=1$ 
contributions to the $e^+ e^-\rightarrow K\bar{K}\pi$ cross
sections~\cite{babarepemkkbarpi} to determine the $V$
part of the $\tau\rightarrow K\bar{K}\pi\nu_\tau$ distribution,
and hence the separation of that distribution into its $V$ and 
$A$ components. The CVC relation allows us to dramatically improve 
the vast majority of the residual-mode contributions to the $V$ 
spectral function.  This is especially helpful in the case of 
contributions from higher-multiplicity modes, whose $\tau$-decay 
distributions lie at higher $s$, increasingly close to the $\tau$ 
kinematic endpoint, and with, as a result, increasingly
reduced statistical precision. 

The main $V$ residual mode for which such a CVC improvement is not 
possible is $K^- K^0$, where the $e^+ e^- \rightarrow K\bar{K}$ 
cross sections contain both $I=0$ and $I=1$ contributions, and it 
is not possible to identify only the $I=1$ component. Fortunately, 
for this channel, BaBar~\cite{babarkkbartau18} has recently published 
a rather precise determination of the unit-normalized 
$\tau\rightarrow K^- K_S\nu_\tau$ number distribution, allowing 
the residual-mode $K^- K^0$ contribution 
to $\rho_{ud;V}(s)$, to be determined directly, without the use of CVC. 

Using CVC and the recent BaBar $\tau$ $K^- K^0$ results, 99.95\% by BF 
of the inclusive $V$ spectral function can be determined directly from 
experiment. The remaining $0.05\%$ represents only 2.4\% by BF of the 
already small sum of residual-mode contributions.
CVC improvements are, of course, impossible for $A$ 
channel residual-mode distributions. This, and the larger relative
role played by residual-mode contributions in the $A$ channel,
are the primary reasons for our focus on the $V$ channel in this paper.

The rest of this section is organized as follows. First, in 
Sec.~\ref{inputs}, we specify the sources of external input employed 
in our update of $\rho_{ud;V}(s)$. Next, in Sec.~\ref{comb}, we outline 
the procedure used for combining the publicly available data from ALEPH 
and OPAL for the dominant $2\pi$ and $4\pi$ exclusive-modes, following
closely that described in Ref.~\cite{KNT18} for combining exclusive-mode 
$e^+ e^-$ cross sections from different experiments. Details of our 
updates of the individual residual exclusive-mode contributions are 
provided in Sec.~\ref{resid}. Finally, the resulting updated version of
$\rho_{ud;V}(s)$, is presented in Sec.~\ref{specfun}.

\subsection{\label{inputs} External input}
As noted above, we employ publicly available results for the non-residual
($\p^-\p^0$, $\p^-3\p^0$ and $\p^-\p^+\p^-\p^0$) exclusive-mode distributions 
and correlations provided by the ALEPH~\cite{ALEPH,ALEPH2,ALEPH13,ALEPH08} 
and OPAL~\cite{OPAL} collaborations. 

ALEPH quotes results in the form of exclusive-mode, BF-normalized
contributions, $dB_X(s)/ds$, to the differential BF distribution 
$dB(s)/ds$. The corresponding contributions to $\rho_{ud;V}(s)$, 
$\rho^X_{ud;V}(s)$, follow from 
\begin{eqnarray}
\rho^X_{ud;V}(s)\, && =\, {\frac{m_\tau^2}{12\pi^2 B_e S_{\rm EW}
\vert V_{ud}\vert^2 w_T(s;m_\t^2)}} \, {\frac{dB_X(s)}{ds}}\nonumber\\
&&=\, {\frac{B_X m_\tau^2}{12\pi^2 B_e S_{\rm EW}
\vert V_{ud}\vert^2 w_T(s;m_\t^2)}} \, {\frac{1}{N_X}}{\frac{dN_X(s)}{ds}}
\label{dBdstorho}\end{eqnarray}
where $B_X$ is the BF for exclusive mode $X$, 
${\frac{1}{N_X}}{\frac{dN_X(s)}{ds}}$ is the corresponding
experimental unit-normalized number distribution, and $B_e$ is the
$\tau^- \rightarrow e^-\nu_\tau \bar{\nu}_e$ BF. ALEPH results for 
$dB_X(s)/ds$ are updated by rescaling to the current value of $B_X$.
Current values of the external parameters, $B_e$, $S_{\rm EW}$, $V_{ud}$
and $m_\tau$ are then used to obtain the updated results for 
$\rho^X_{ud;V}(s)$.

OPAL quotes results in the form of the exclusive-mode contributions 
$\rho^X_{ud;V}(s)$. These were obtained from the experimentally measured 
unit-normalized number distributions via Eq.~(\ref{dBdstorho}), using
then-current values of the exclusive-mode BFs, $B_X$, and the external 
inputs $B_e$, $S_{\rm EW}$, $V_{ud}$ and $m_\tau$. The underlying 
unit-normalized distributions are reconstituted using the values for
the exclusive-mode BFs and external inputs quoted by OPAL, and 
converted to equivalent updated versions of the $\rho^X_{ud;V}(s)$ 
using current values for these inputs.

We employ the following values for the external parameters appearing 
in Eq.~(\ref{dBdstorho}): for $B_e$, the lepton-universality-improved HFLAV 
2019~\cite{hflav2019} result $B_e=0.17814(22)$; for $m_\tau$ and
$\vert V_{ud}\vert$, the PDG 2020~\cite{pdg2020} results 
$m_\tau =1.77686(12)$ GeV and $\vert V_{ud}\vert=0.97370(14)$;
and, for $S_{\rm EW}$, the result $S_{\rm EW}=1.0201(3)$~\cite{erlersew}.

For the exclusive-mode BFs and the correlations between them we employ 
HFLAV 2019~\cite{hflav2019} results. Note that HFLAV quotes a result
for the $\pi^-\pi^+\pi^-\pi^0$ BF which excludes $K^0$ contributions
but not the small ``wrong-current'' $A$ $\pi^-\pi^0\omega (\rightarrow
\pi^+\pi^-)$ contribution. The $V$ part of the $\pi^-\pi^+\pi^-\pi^0$
BF is obtained by removing this $A$ ``contamination'' using the HFLAV
$\pi^-\pi^0\omega$ BF and 2020 PDG~\cite{pdg2020} result for the
IB $\omega\rightarrow\pi^-\pi^+$ BF. Similar wrong-current corrections
are made to the correlations between the BFs.

Specifics of the experimental inputs used in the determination of the 
residual-mode contributions to $\rho_{ud;V}(s)$ are detailed in
Sec.~\ref{resid} below.

\begin{boldmath}
\subsection{\label{comb} Combining the ALEPH and OPAL $2\p$ and $4\p$ data}
\end{boldmath}
We begin with the updated versions of the ALEPH and OPAL exclusive-mode
distributions, $\rho^X_{ud;V}(s)$, with $X=\p^-\p^0$, $\p^-3\p^0$ and 
$\p^-\p^+\p^-\p^0$, obtained as outlined in the previous subsection.
Ideally, we would like to combine the results for each of these exclusive 
modes separately, first combining the ALEPH and OPAL unit-normalized number 
distributions (which are independent of the BFs) and then multiplying the 
resulting combined exclusive distributions by the corresponding BFs. It
turns out, however, that this is not possible. The reason is that the 
correlation matrices for the $\p^-3\p^0$ and $\p^-\p^+\p^-\p^0$ 
distributions for both experiments have zero eigenvalues, \ie, 100\% 
correlations between different bins, and hence are not invertible. This
prevents us from combining the ALEPH and OPAL data for the individual
$4\pi$ modes in the manner described below. If, however, we first sum 
the contributions from all three modes, we find that the correlation matrices 
for the resulting three-mode-sums are well behaved for both ALEPH and OPAL. 
We thus combine the ALEPH and OPAL exclusive-mode results by first summing,
for each experiment separately, the contribution to $\rho_{ud;V}(s)$ and the 
corresponding covariance matrices from $\p^-\p^0$, $\p^-3\p^0$ and 
$\p^-\p^+\p^-\p^0$, using updated versions of the exclusive-mode BFs,
and then combining those results using the method outlined below.

Following Ref.~\cite{KNT18}, we choose a number of {\it clusters}, distributed 
over the interval $0<s\le m_\t^2$. We assign a number of consecutive
ALEPH and OPAL data points to each cluster $m$, $m=1,\dots,N_{\rm cl}$,
with $N_{\rm cl}$ the total number of clusters; $N_m$ will be the total 
number of data points in cluster $m$. If the collective ALEPH and OPAL 
data points are parametrized by pairs $(s_{i},d_{i})$, where $d_{i}$ is 
the ALEPH or OPAL data point for the spectral function assigned 
to the $s$-value $s_{i}$, we define weighted cluster averages
\begin{equation}
\label{clusters}
s^{(m)}=\sum_{i\in m}\frac{s_{i}}{\s^2_{i}}\Bigg/\sum_{i\in m}\frac{1}{\s^2_{i}}\ ,
\end{equation}
where the sum is over all data points in cluster $m$ and $\s^2_i$ is the 
variance of $d_i$, \ie, the $\s^2_i$ are the diagonal elements of the 
covariance matrix $C_{ij}$ for the spectral-function data points $d_i$.
The set of $s^{(m)}$ then constitutes the values of $s$ at which the 
combined spectral function $\r^{(m)}$ will be defined.

The values of $\r^{(m)}$ will be determined by linear interpolation, 
minimizing
\begin{equation}
\label{chi2}
\c^2(\r)=\sum_{i=1}^{N}\sum_{j=1}^{N}\left(
d_{i}-R(s_i;\r)\right)C^{-1}_{i j}\left(d_{j}-R(s_j;\r)\right)\ ,
\end{equation}
where $N=\sum_{m=1}^{N_{\rm cl}}N_m$ is the total number of (ALEPH and OPAL) 
data points, and the piece-wise linear function $R(s;\r)$ is defined by
\begin{equation}
\label{defR}
R(s;\r)=\r^{(m)}+\frac{s-s^{(m)}}{s^{(m+1)}-s^{(m)}}\left(\r^{(m+1)}-\r^{(m)}\right)\ ,\quad s^{(m)}\le s\le s^{(m+1)}\ ,\quad 1\le m<N_{\rm cl}\ ,
\end{equation}
where $\r$ is the vector of fit parameters $\r^{(m)}$, $m=1,\dots,N_{\rm cl}$.
At the boundaries, we extrapolate:
\begin{eqnarray}
\label{defR2}
    R(s;\r)&=&\r^{(N_{\rm cl}-1)}+\frac{s-s^{(N_{\rm cl}-1)}}{s^{(N_{\rm cl})}-s^{(N_{\rm cl}-1)}}\left(\r^{(N_{\rm cl})}-\r^{(N_{\rm cl}-1)}\right)\ ,\qquad s\ge s^{(N_{\rm cl})}\ ,
\\
R(s;\r)&=&
\r^{(1)}+\frac{s-s^{(1)}}{s^{(2)}-s^{(1)}}\left(\r^{(2)}-\r^{(1)}\right)\ ,\qquad\qquad\qquad\qquad\ s\le s^{(1)}\ .\nonumber
\end{eqnarray}
Minimizing $\c^2(\r)$ yields the linear equations
\begin{equation}
\label{minchi2}
\sum_{i=1}^{N}\sum_{j=1}^{N}\left(
d_{i}-R(s_i;\r)\right)C^{-1}_{i j}\,\frac{\partial R(s_j;\r)}{\partial\r^{(m)}}=0\ ,\qquad 1\le m\le N_{\rm cl}\ ,
\end{equation}
which can be solved for the $\r^{(m)}$, with the cluster covariance 
matrix $\cc_{mn}$ given by
\begin{equation}
\label{clcov}
\cc_{mn}^{-1}=\sum_{i=1}^{N}\sum_{j=1}^{N}\frac{\partial R(s_i;\r)}{\partial\r^{(m)}}\,C^{-1}_{i j}\,\frac{\partial R(s_j;\r)}{\partial\r^{(n)}}\ .
\end{equation}

The procedure for combining exclusive spectral functions followed 
in Ref.~\cite{KNT18} is more complicated than the one outlined above. 
The inclusion of uncertainties in the BFs in the covariance matrices 
can lead to a bias in the fit \cite{Abias}, and the method employed 
in Ref.~\cite{KNT18} adjusts for this bias \cite{NNPDF}. However, for this 
to work, we would need to combine each channel separately, because 
multiplication by the exclusive-mode BF is needed in each channel 
to turn the normalized distribution into the corresponding contribution 
to $\rho_{ud;V}(s)$. This path is not available to us, because, as
explained above, only the sum of the $\p^-\p^0$, $\p^-3\p^0$ and 
$\p^-\p^+\p^-\p^0$ spectral distributions can be combined. This sum 
incorporates three different branching fractions, one for each exclusive 
channel. The effect of the BF uncertainties is, however, very 
small numerically. We have checked that their inclusion changes central 
values of the combined three-mode contribution to $\rho_{ud;V}(s)$
by less than 0.5\%, while the errors on this contribution are about one 
percent larger with than without inclusion of BF uncertainties, and the 
cluster values $s^{(m)}$ are essentially unaffected. It is thus safe to 
ignore potential bias issues associated with the incorporation of BF
uncertainties. The same is true for the impact of the uncertainties 
in the external normalizing factors $V_{ud}$, $S_{\rm EW}$, and $B_e$.
The errors on $V_{ud}$ and $S_{\rm EW}$ are completely negligible, 
while the error on $B_e$, at about 0.1\% is small enough that it does 
not lead to a discernible bias.

In order to obtain an optimal combined data set, one needs to choose the 
clusters judiciously. Clearly, the maximum number of clusters one can 
choose is equal to the sum of the total number of ALEPH and OPAL data 
points. However, such a choice is not useful. One would find a $\c^2$ 
per degree of freedom (dof) much smaller than one, but this is not
what one expects if one averages two independent experiments. Since both 
ALEPH and OPAL measured the spectral function over the same range in $s$, 
the number of clusters should be chosen not larger than the number
of data points in each of the experiments, and, given that these two 
experiments are independent,\footnote{Correlations introduced by the 
use of the same BFs for both experiments are negligibly small.} one expects a
$\c^2/$dof of order one. The goal is thus to choose a set of clusters not 
larger than the number of data points in either experiment that leads
to a value of $\c^2(\r)/\mbox{dof}\approx 1$. Of course, narrower 
clusters should be used in regions where the spectral function changes 
rapidly, such as around the $\r$-meson peak.

In addition to the ``global'' $\c^2(\r)$ function defined in Eq.~(\ref{chi2}), 
we have also looked at $\c^2_{(m)}$, the ``local'' $\c^2$ function for
each cluster, since the local $\chi^2_{(m)}$ values may reveal discrepancies 
in the data sets that are hidden in the global $\chi^2$. The local 
$\c^2_{(m)}$ is defined as in Eq.~(\ref{chi2}), but with both the data 
points $(s_i,d_i)$ and the data covariance matrix restricted to those 
data points contained in cluster $m$. We then evaluate all $\c^2_{(m)}$ 
on the solution $\r^{(m)}$, $m=1,\dots,N_{\rm cl}$, obtained by minimizing 
the global $\c^2(\r)$. Clearly, the global $\c^2(\r)$ is not equal
to the sum over all clusters of the local $\c^2_{(m)}$, because the full 
data covariance matrix $C$ contains entries correlating data points in 
different clusters. If, for a cluster $k$, $\c^2_{(k)}/$dof$>1$, this 
indicates a fluctuation or a local discrepancy between the ALEPH and OPAL 
data.

At this stage, we will have obtained a partially-inclusive combined 
spectral function and associated covariance matrix for the sum of the 
$\p^-\p^0$, $\p^-3\p^0$ and $\p^-\p^+\p^-\p^0$
modes. We still need to add the residual-mode contributions to obtain our
final, updated version of $\rho_{ud;V}(s)$. The determination of
the residual-mode contributions is detailed in the next subsection. 

\vskip0.7cm

\subsection{\label{resid} Residual Mode Updates}
In this section we provide details of the input used to update the 
residual exclusive-mode contributions to $\rho_{ud;V}(s)$, \ie, all 
modes other than $\p^-\p^0$, $\p^-3\p^0$ or $\p^-\p^+\p^-\p^0$. The modes 
considered in this work are (i) those included in both the OPAL and 
ALEPH analyses, $\pi^- \omega (\rightarrow {\rm \mbox{non-}3}\pi)$, $K^- K^0$,
$\eta \pi^- \pi^0$, $K\bar{K}\pi$, $3\pi^- 2\pi^+ \pi^0$, and
$2\pi^- \pi^+ 3\pi^0$, (ii) those included in the ALEPH analysis but not 
the OPAL analysis, $(3\pi )^- \omega(\rightarrow {\rm \mbox{non-}3}\pi)$
and $K\bar{K}\pi\pi$, and (iii) small additional $\pi^- 5\pi^0$ and 
$\eta \omega\pi +\eta 4\pi$ contributions inferrable from the 
corresponding $e^+ e^-$ cross sections using CVC, and not included in 
either of the OPAL or ALEPH analyses. 

Note that, where the $e^+ e^-\rightarrow hadrons$ cross sections used to 
infer, via CVC, the corresponding contributions to $\rho_{ud;V}(s)$, 
are given in dressed form in the original publications, these
have been corrected for vacuum polarization effects to obtain
the corresponding bare cross sections required as input to the
CVC relation.   Statistical and systematic errors on the
cross sections are those reported in the relevant references. 
Additional information, if any, provided by the collaborations
is specified below.

All results for $\tau$ exclusive-mode BFs quoted
below are obtained using basis-mode BF and correlation 
information from the 2019 HFLAV compilation~\cite{hflav2019}.

We now turn to a more detailed discussion of the determination of
the residual exclusive-mode contributions to $\rho_{ud;V}(s)$. 
The discussion is organized mode by mode, in the order of decreasing 
residual-mode BF. Readers interested only in the final result
for the inclusive spectral function may skip these details and
jump directly to Sec.~\ref{specfun} below.

\subsubsection{\label{piomeganon3pi} The $\pi^- \omega (\rightarrow \mbox{non-}3\pi)$ contribution}

The $\pi^- \omega (\rightarrow {\rm \mbox{non-}3}\pi)$ contribution to 
$\rho_{ud;V}(s)$ is obtained using CVC, BaBar results~\cite{babaromegapi17} 
for the $e^+ e^-\rightarrow \pi^0\omega$ cross sections, and the 2020 PDG 
value, $0.107(6)$, for the $\omega \rightarrow {\rm \mbox{non-}3}\pi$ BF. The
BaBar cross sections are in good agreement with, and have significantly 
smaller errors than, those reported by SND~\cite{sndomegapi16}. 
The BaBar results produce a CVC prediction of $0.0188(19)$ 
for the $\tau\rightarrow\pi^-\omega \nu_\tau$ BF, in excellent 
agreement with the HFLAV 2019 result, $0.01955(65)$. The contribution 
to $\rho_{ud;V}(s)$ implied by the BaBar cross sections has been rescaled 
by the ratio of the HFLAV to the CVC BF to normalize it to the HFLAV 
2019 $\tau$ BF. The BF corresponding to the resulting 
$\pi^- \omega (\rightarrow {\rm \mbox{non-}3}\pi)$ contribution to 
$\rho_{ud;V}(s)$ is $0.00209(14)$.

\subsubsection{\label{etapipi} The $\eta \pi^-\pi^0$ contribution}

The $e^+ e^-\rightarrow\eta \pi^+\pi^-$ cross sections have been
measured by SND~\cite{sndetapipi1518,sndetapipi15182}, 
BaBar~\cite{babaretapipi18,babaretapipi182}
and CMD-3~\cite{cmd3etapipi19}, with the results from all three 
collaborations in excellent agreement (see, for example, Figure 7 
of Ref.~\cite{cmd3etapipi19}). Since CMD-3 has provided us with the 
corresponding covariances \cite{cmd3etapipicovs}, we employ the 
CMD-3~\cite{cmd3etapipi19} cross section data as input to our CVC 
determination of the $\eta\pi^-\pi^0$ contribution to $\rho_{ud;V}(s)$. 
As noted by CMD-3, the SND, BaBar and CMD-3 cross sections produce CVC
predictions for the $\tau\rightarrow\eta\pi^-\pi^0\nu_\tau$ BF,
$0.00156(11)$, $0.00163(8)$ and $0.00168(13)$, respectively,
which are in good agreement, but which lie between $1.3$ and $2.3$ 
$\sigma$ high compared to the corresponding HFLAV 2019 result, 
$0.001386(72)$. Since the HFLAV 2019 $\tau$ average is strongly 
dominated by a single (Belle~\cite{belleetapipi09}) experiment,
we have normalized the residual-mode $\eta\pi^-\pi^0$ contribution
using a BF value, $0.00153(12)$ obtained by averaging, with PDG-style 
error inflation, the results of the three CVC predictions and
the HFLAV 2019 result. 

\subsubsection{\label{kmk0} The $K^- K^0$ contribution}

The $K^- K^0$ contribution to $\rho_{ud;V}(s)$ is obtained using 
BaBar results~\cite{babarkkbartau18} for the unit-normalized 
$\tau\rightarrow K^- K_s\nu_\tau$ number distribution, normalized to 
the HFLAV 2019 value of the $\tau\rightarrow K^- K^0\nu_\tau$
BF, $0.001483(34)$. 

\subsubsection{\label{kkbarpi} The $K\bar{K}\pi$ contribution}

Determining the $K\bar{K}\pi$ contribution to $\rho_{ud;V}(s)$ is 
less straightforward since the measured distribution in 
$\tau\rightarrow K\bar{K}\pi\nu_\tau$ is a sum of $V$ and $A$ 
contributions, while the $e^+ e^-\rightarrow K\bar{K}\pi$ cross sections 
are sums of $I=0$ and $I=1$ contributions. The $V$ and $A$ contributions 
to $\tau\rightarrow K\bar{K}\pi\nu_\tau$ cannot be separated without an 
angular analysis, which has not been carried out to date. 
BaBar~\cite{babarepemkkbarpi}, however, has succeeded in using a Dalitz-plot 
analysis to separate the $I=0$ and $I=1$ parts of the 
$e^+ e^-\rightarrow KK^*\rightarrow K\bar{K}\pi$ cross sections, which, with 
the smaller $e^+ e^-\rightarrow\pi^0\phi\rightarrow \pi^0 K\bar{K}$
contributions, dominate the $e^+ e^-\rightarrow K\bar{K}\pi$ cross section
at CM energies below $m_\tau$. ALEPH~\cite{ALEPH08} has previously used the 
$I=1$ $V$ cross sections extracted in this analysis, together with CVC, to 
determine the $V$ component of the $\tau\rightarrow K\bar{K}\pi\nu_\tau$
BF. Following the ALEPH strategy, we obtain the sum of the contributions from 
the three $K\bar{K}\pi$ states to $\rho_{ud;V}(s)$ using CVC, the $I=1$ 
$e^+ e^-\rightarrow KK^*\rightarrow K\bar{K}\pi$ and
$e^+ e^-\rightarrow\pi^0\phi$ cross sections measured by 
BaBar~\cite{babarepemkkbarpi}, standard vacuum-polarization corrections 
to convert these to the corresponding bare cross sections,
and the 2020 PDG value for the $\phi\rightarrow K\bar{K}$ 
BF.\footnote{For reference, this produces a CVC expectation of 
$0.00073(9)$ for the $V$ part of the sum of the three 
$\tau\rightarrow K\bar{K}\pi\nu_\tau$ BFs. This represents 
$16.4\pm 2.2\%$ of the $0.00444(26)$ HFLAV 2019 result 
for the 3-mode $V+A$ BF sum.} 

\subsubsection{\label{6pi} The $6\pi$ contributions}

The sum of the three $6\pi$ mode contributions to $\rho_{ud;V}(s)$ is 
obtained using CVC in conjunction with the measured $e^+ e^-\rightarrow 6\pi$ 
and $\pi^+\pi^-\pi^0\eta$ cross sections. The $\pi^+\pi^-\pi^0\eta$ cross 
sections are required because the IB $\eta\rightarrow \pi^+\pi^-\pi^0$ 
and $3\pi^0$ decays cause the $G$-parity negative, $I=0$ $\pi^+\pi^-\pi^0\eta$ 
state to also populate the experimental $2\pi^- 2\pi^+ 2\pi^0$ and
$\pi^-\pi^+ 4\pi^0$ distributions. These ``wrong current'' contributions 
must be removed in order to obtain the $I=1$ components of these 
distributions to which the CVC relation may be applied. We employ 
BaBar~\cite{babarsigma6pi06} and CMD-3~\cite{cmd3sigma3pim3pip13} results 
for the $e^+ e^-\rightarrow 3\pi^- 3\pi^+$ cross sections, 
BaBar~\cite{babarsigma6pi06} results for the unsubtracted
$e^+ e^-\rightarrow 2\pi^- 2\pi^+ 2\pi^0$ cross sections, the
preliminary SND results reported in Ref.~\cite{sndsigmapimpip4pi019}
for the unsubtracted $e^+ e^-\rightarrow \pi^-\pi^+ 4\pi^0$ cross sections, 
and CMD-3~\cite{cmd3sigmaeta3pi17} and SND~\cite{sndsigmaeta3pi19}
results for the $e^+ e^-\rightarrow \pi^+\pi^-\pi^0\eta$ cross sections.
The three-mode $I=1$ $6\pi$ cross section sum produces a CVC prediction
of $0.000280(35)$ for the sum of the $V$ components of the 
$\tau\rightarrow 3\pi^- 2\pi^+\pi^0\nu_\tau$, 
$\tau\rightarrow 2\pi^- \pi^+ 3\pi^0\nu_\tau$ and
$\tau\rightarrow \pi^- 5\pi^0\nu_\tau$ BFs.\footnote{This cannot be compared
to the corresponding HFLAV 2019 version for this three-mode sum since the
BF for $\tau\rightarrow \pi^- 5\pi^0\nu_\tau$ has not yet been measured. 
The HFLAV 2019 results for the remaining two $\tau\rightarrow 6\pi \nu_\tau$ 
BFs, in addition, have non-negligible $A$ contributions which must be 
subtracted in order to identify the purely $V$ contributions. The 
``wrong current'' $A$ contributions to the G-parity-positive $6\pi$ 
states are the result of IB $\eta\rightarrow 3\pi$ decays, which cause
$A$ $2\pi^-\pi^+\eta$ and $\pi^- 2\pi^0\eta$ states to populate
the experimental $3\pi^- 2\pi^+ \pi^0$ and $2\pi^- \pi^+ 3\pi^0$
distributions. Using HFLAV 2019 results for the unsubtracted 
$6\pi$ BFs (excluding $K^0$ contributions) and the two $3\pi \eta$ BFs, 
together with 2020 PDG values for the $\eta\rightarrow \pi^+\pi^-\pi^0$ and 
$\eta\rightarrow 3\pi^0$ BFs, one finds for the $V$ contributions to the 
$\tau\rightarrow 3\pi^- 2\pi^+ \pi^0\nu_\tau$ and
$\tau\rightarrow 2\pi^- \pi^+ 3\pi^0\nu_\tau$ BFs the results $0.000113(10)$ 
and $0.000077(28)$, respectively. The CVC prediction for the
3-mode $\tau\rightarrow 6\pi\nu_\tau$ $V$ BF sum thus corresponds
to a $V$ contribution of $0.000090(46)$ to the BF of 
$\tau\rightarrow \pi^- 5\pi^0\nu_\tau$.}

\subsubsection{\label{kkbar2pi} The $K\bar{K}\pi\pi$ contributions}

No direct experimental determination of the $K\bar{K}\pi\pi$ 
contributions to $\rho_{ud;V}(s)$ is currently available. Even were 
$\tau\rightarrow K\bar{K}\pi\pi \nu_\tau$ distribution results publicly 
available, no obvious strategy exists for splitting this distribution 
into its separate $V$ and $A$ parts.  The BF situation is, moreover,
incomplete for $\tau\rightarrow K\bar{K}\pi\pi \nu_\tau$ decays, with 
HFLAV listing BFs for only two of the five possible $K\bar{K}\pi\pi$ $\tau$ 
modes~\cite{hflav2019}. 

The experimental situation is more complete for 
$e^+ e^-\rightarrow K\bar{K}\pi\pi$, with cross sections available 
for all six $K\bar{K}\pi\pi$ final states. These cross sections are,
however, at-present-unknown admixtures of $I=0$ and $I=1$ contributions,
with no known method for separating the $I=0$ and $I=1$ components.
This precludes a CVC determination of the $K\bar{K}\pi\pi$ contribution
to $\rho_{ud;V}(s)$. The unseparated $I=0+1$ cross sections, and the 
resulting full $I=0+1$ six-mode sum, are, however, rather accurately 
known. In what follows we rely on the results for this sum obtained 
in Ref.~\cite{KNT18} as part of the recent dispersive determination of 
the hadronic vacuum polarization contribution to the anomalous 
magnetic moment of the muon, and provided to us by the authors~\cite{thanksAK}.

In Ref.~\cite{ALEPH,ALEPH2,ALEPH13,ALEPH08}, ALEPH employed a maximally conservative 
approach to the $K\bar{K}\pi\pi$ contribution to $\rho_{ud;V}(s)$, 
assigning $50\pm 50\%$ of the $V+A$ $\tau\rightarrow K\bar{K}\pi\pi \nu_\tau$ 
distribution to $\rho_{ud;V}(s)$. With current HFLAV 2019 values, 
the sum of the two currently known $\tau\rightarrow K\bar{K}\pi\pi\nu_\tau$ 
BFs (those for $\tau\rightarrow \pi^-\pi^0 K^0\bar{K}^0\nu_\tau$ and
$\tau\rightarrow \pi^-\pi^0 K^- K^+ \nu_\tau$) is $0.0004154(1207)$.
The ALEPH choice would thus correspond to a $V$ contribution to the
all-modes $\tau\rightarrow K\bar{K}\pi\pi\nu_\tau$ BF sum, from these
two modes only, of $0.000208(60)(208)$, where the first error is 50\% of 
the error on the HFLAV $V+A$ sum and the second represents the assigned 
100\% uncertainty on the separation of the $V+A$ sum into $V$ and $A$ 
components. 

In contrast, if one makes the analogous maximally conservative 
assessment and assigns $50\pm 50\%$ of the six-mode sum of $I=0+1$ 
$e^+ e^-\rightarrow K\bar{K}\pi\pi$ cross sections to $I=1$, the CVC
relation yields a sum of the contributions from all $K\bar{K}\pi\pi$ modes 
to $\rho_{ud;V}(s)$ which corresponds to an all-modes $V$ 
$\tau\rightarrow K\bar{K}\pi\pi\nu_\tau$ BF sum of $0.000154(5)(154)$,
where the second error reflects the assigned, maximally conservative 
100\% $I=0/1$ separation uncertainty. Since this constraint on the 
$V$ $K\bar{K}\pi\pi$ contribution is stronger than that resulting 
from the alternate maximally conservative assessment based on the 
two measured $\tau\rightarrow K\bar{K}\pi\pi\nu_\tau$ BFs, we employ 
the CVC assessment and assign as the $K\bar{K}\pi\pi$ contribution
to $\rho_{ud;V}(s)$, $50\pm 50\%$ of the result obtained by applying
the CVC relation to the full six-mode $I=0+1$ $e^+ e^-\rightarrow
K\bar{K}\pi\pi$ cross section sum. The CVC determination
has the additional advantage that it includes contributions from
all $K\bar{K}\pi\pi$ modes, even those for which the corresponding 
$\tau$ BFs are currently unknown.

\subsubsection{\label{3piomeganon3pi} The $(3\pi )^-\omega (\rightarrow {\rm \mbox{non-}3}\pi)$ contributions}

Contributions to $\rho_{ud;V}(s)$ from the last of the $V$ residual modes 
considered by ALEPH, $(3\pi )^-\omega (\rightarrow {\rm \mbox{non-}3}\pi)$, 
can be obtained from the corresponding $(3\pi )^-\omega (\rightarrow 3\pi )$
contributions using the known values of the $\omega\rightarrow 3\pi$ and
$\omega\rightarrow {\rm \mbox{non-}3}\pi$ BFs. While the relevant $\tau$-decay 
distributions have not yet been measured, BaBar~\cite{babarsigma6pi06} has 
determined both the
$e^+e^-\rightarrow\pi^-\pi^+\pi^0 \omega (\rightarrow \pi^-\pi^+\pi^0 )$
contribution to the $e^+e^-\rightarrow 2\pi^- 2\pi^+ 2\pi^0$ cross sections
and the ``wrong-current'' $I=0$ 
$\omega (\rightarrow \pi^-\pi^+\pi^0 ) \eta (\rightarrow \pi^-\pi^+\pi^0 )$
component of that contribution \cite{thanksSolodov}. We subtract this wrong-current
contribution to obtain the $I=1$ contributions to the 
$\pi^-\pi^+\pi^0 \omega (\rightarrow \pi^-\pi^+\pi^0 )$ cross sections, use 
2020 PDG versions of the $\omega$-decay BFs to obtain the corresponding $I=1$ 
contributions to the 
$\pi^-\pi^+\pi^0 \omega (\rightarrow {\rm \mbox{non-}3}\pi )$
cross sections, and the CVC relation to determine the corresponding
contributions to $\rho_{ud;V}(s)$. 

$3\pi^0 \omega (\rightarrow 3\pi)$ 
contributions are also, in principle, present in the 
$e^+ e^-\rightarrow \pi^-\pi^+ 4\pi^0$ cross sections. The preliminary 
SND results for the latter~\cite{sndsigmapimpip4pi019} do not include 
an assessment of the $e^+ e^-\rightarrow 3\pi^0 \omega (\rightarrow 3\pi)$
substate contribution. The following argument, however, shows these 
contributions, though not measured, must be small enough to be ignored 
in our CVC determination of the residual-mode 
$(3\pi )^-\omega (\rightarrow {\rm \mbox{non-}3}\pi)$ contribution to 
$\rho_{ud;V}(s)$. 

Explicitly, HFLAV 2019 results for the BFs of
the $\tau\rightarrow 2\pi^-\pi^+\omega\nu_\tau$ and 
$\tau\rightarrow \pi^-2\pi^0\omega\nu_\tau$ modes yield a value of
$0.000155(18)$ for the $\tau\rightarrow (3\pi )^-\omega \nu_\tau$ BF 
sum. Applying the CVC relation to the $I=1$ component of the BaBar
$e^+e^-\rightarrow\pi^-\pi^+\pi^0 \omega (\rightarrow \pi^-\pi^+\pi^0 )$
cross sections, one finds a CVC prediction for the contribution to the
two-mode $\tau$ BF sum of $0.000172(25)$, compatible within errors with
the full 2-mode $\tau\rightarrow (3\pi )^-\omega$ HFLAV BF result. We 
conclude that $I=1$ $3\pi^0\omega (\rightarrow \pi^-\pi^+\pi^0)$ 
contributions to the $e^+ e^-\rightarrow \pi^-\pi^+ 4\pi^0$ cross 
sections, which would produce a further increase in the CVC prediction 
for the full 2-mode $\tau$ BF sum, must be numerically small. We
thus determine the $(3\pi )^-\omega (\rightarrow {\rm \mbox{non-}3}\pi)$
contribution to $\rho_{ud;V}(s)$ by (i) applying the CVC relation to the
$I=1$ part of the BaBar results for the 
$e^+e^-\rightarrow\pi^-\pi^+\pi^0 \omega (\rightarrow \pi^-\pi^+\pi^0 )$
cross sections, (ii) dividing those result by the $\omega\rightarrow 3\pi$ 
BF~\cite{pdg2020} to obtain the corresponding all-modes 
$\tau\rightarrow (3\pi )^-\omega\nu_\tau$ contribution to $\rho_{ud;V}(s)$, 
(iii) rescaling this result to normalize it to the HFLAV 2019 result for the
$\tau\rightarrow (3\pi )^-\omega\nu_\tau$ BF, and (iv) multiplying this
result by the $\omega\rightarrow {\rm \mbox{non-}3}\pi$ BF~\cite{pdg2020} to
obtain the final $(3\pi )^-\omega (\rightarrow {\rm \mbox{non-}3}\pi)$
contribution.

\subsubsection{\label{pietaomegaeta4pi} The $\pi^-\eta\omega (\rightarrow {\rm \mbox{non-}3}\pi )$ and $\eta 4\pi$ contributions}

The final $V$ residual-mode contribution we consider is that produced by the
$\pi^- \eta\omega (\rightarrow {\rm \mbox{non-}3}\pi )$ and $\eta (4\pi )^-$ 
modes. This is evaluated using CVC and BaBar results for the   
$e^+ e^-\rightarrow \pi^0\eta \omega$~\cite{babarepemetapimpip2pi018},
$e^+ e^- \rightarrow \eta 2\pi^- 2\pi^+$~\cite{babarepemeta2pim2pip07} and 
$e^+ e^- \rightarrow \eta \pi^- \pi^+ 2\pi^0$~\cite{babarepemetapimpip2pi018} 
cross sections. SND results with significantly larger errors, also exist for 
the $e^+ e^-\rightarrow \pi^0\eta \omega$ cross 
sections~\cite{sndepemetaetaomegapi16}. 

The results of 
Ref.~\cite{babarepemetapimpip2pi018} show that the contribution from 
$e^+ e^-\rightarrow \pi^0 \eta \omega (\rightarrow 3\pi )$
saturates the $e^+ e^- \rightarrow \eta \pi^- \pi^+ 2\pi^0$
cross section below $s=m_\tau^2$. We thus take the sum of 
$e^+ e^- \rightarrow \eta 2\pi^- 2\pi^+$ and 
$e^+ e^-\rightarrow \pi^0\eta \omega$ cross sections as input to
the CVC relation, obtaining, as a result, the sum of 
$\eta (4\pi )^-$ and $\pi^-\eta\omega (\rightarrow {\rm \mbox{non-}3}\pi )$
contributions to $\rho_{ud;V}(s)$, which we identify by the
short-hand label $\eta\omega\pi\eta 4\pi$ in what follows. 

The resulting contribution to $\rho_{ud;V}(s)$ corresponds
to a very small, $0.0000017(2)$, result for the associated
$\tau$ BF sum. This provides further support for the expectation 
that contributions from additional higher-multiplicity $V$ residual 
modes not included in the present analysis will be entirely
numerically negligible in the region below $s=m_\tau^2$.

%%%%%%%%%%%%%%%%%%%
\begin{figure}[t]
%\vspace*{4ex}
\begin{center}
\includegraphics*[width=7.3cm]{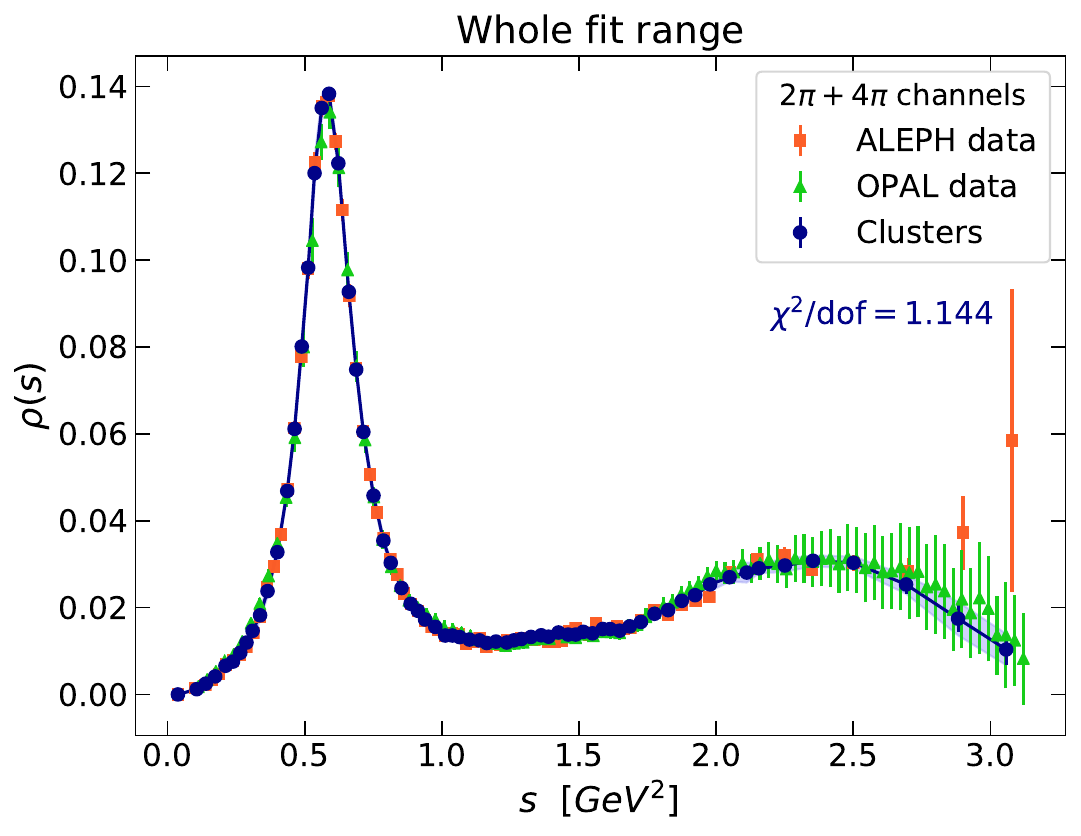}
\hspace{0.2cm}
\includegraphics*[width=7.3cm]{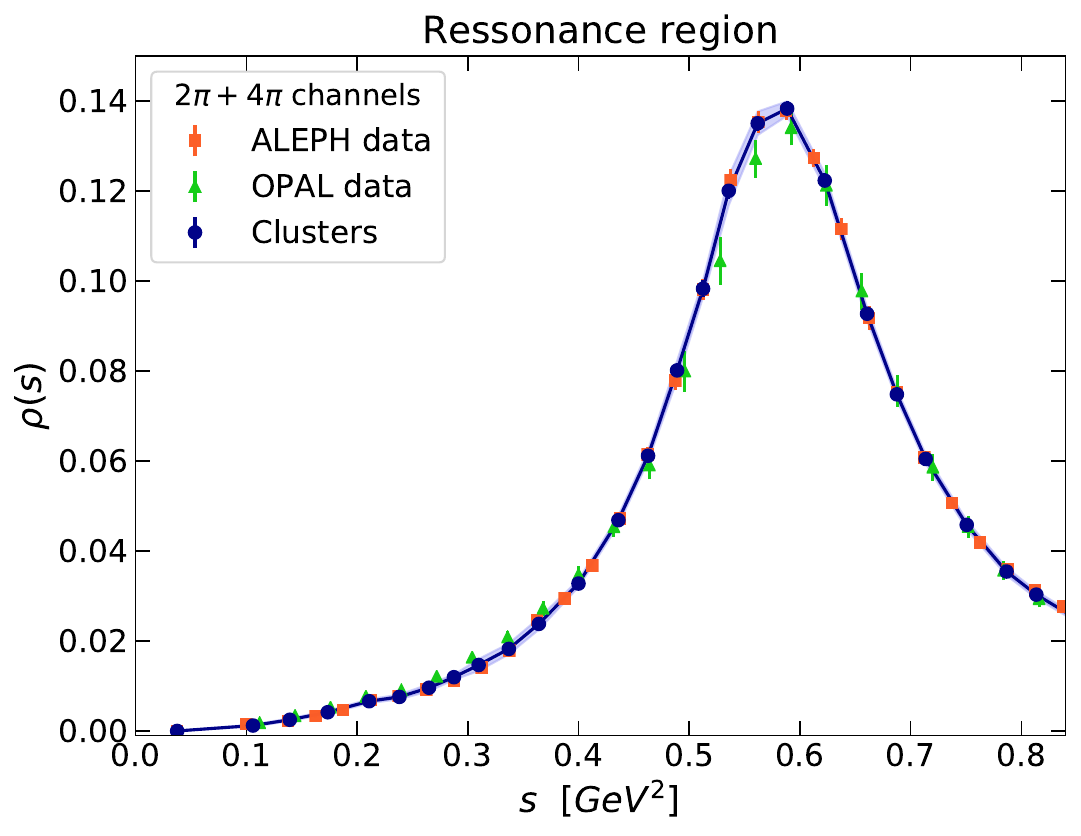}
\vspace{1cm}
\includegraphics*[width=7.3cm]{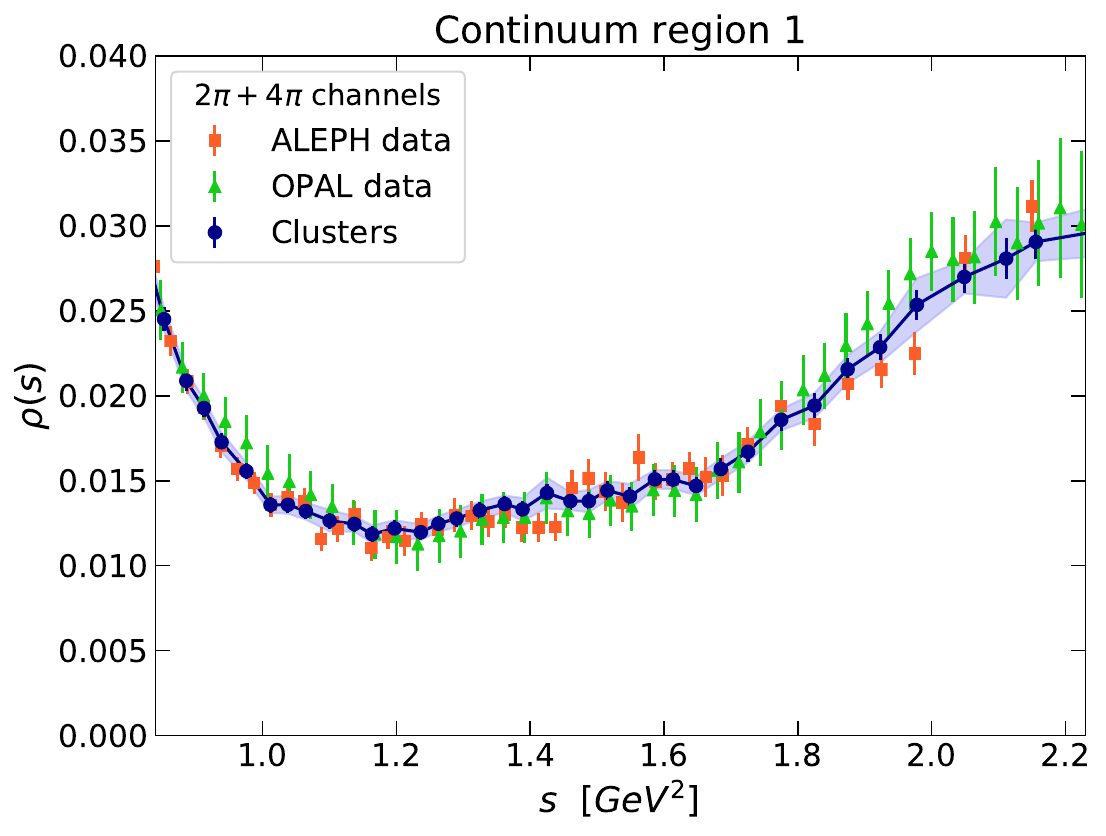}
\hspace{0.2cm}
\includegraphics*[width=7.3cm]{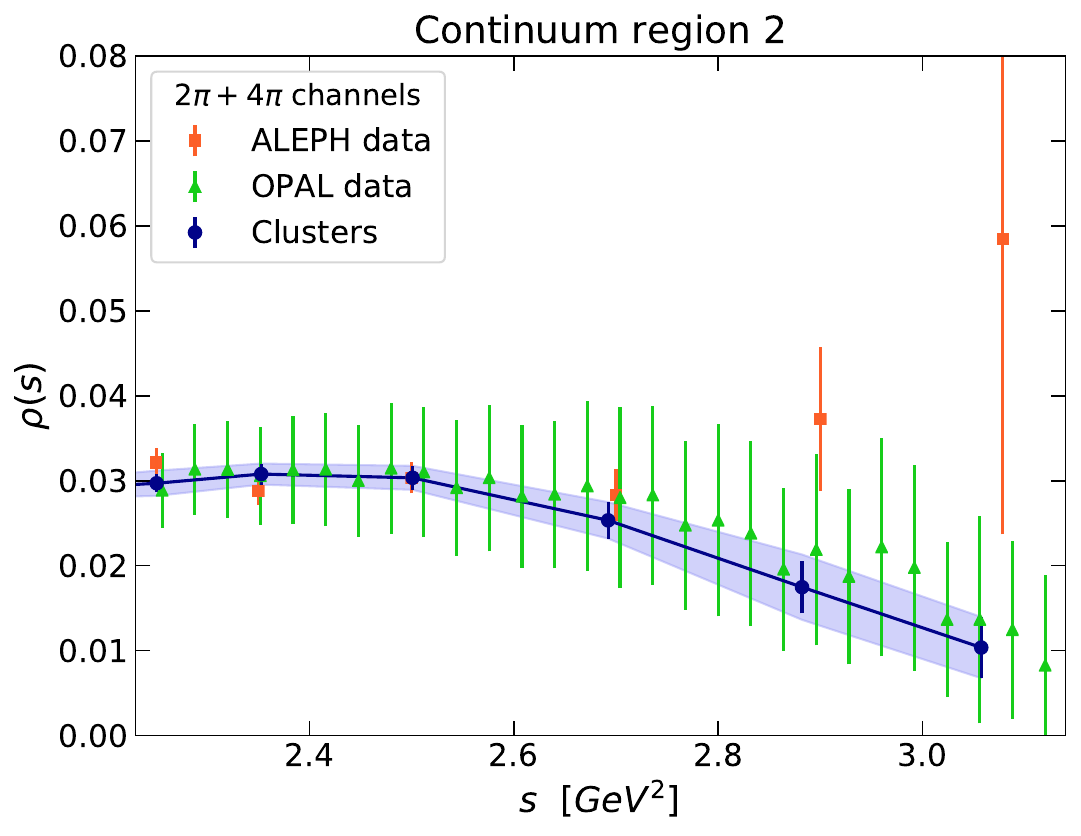}
\end{center}
\vspace*{-10ex}
\begin{quotation}
\floatcaption{figcl1}{{\it Result of the fit for the 68-cluster 
combination of the $2\p+4\p$ channels. The error bars represents 
non-inflated errors while inflated errors are represented by the 
blue band. }}
\end{quotation}
\vspace*{-4ex}
\end{figure}
%%%%%%%%%%%%%%%%%%%

\begin{boldmath}
\subsection{\label{specfun} The inclusive $V$ non-strange spectral function.}
\end{boldmath}
The main decision to be made when combining the data into clusters is 
the choice of the clusters themselves. One possibility is the basis of 
the strategy used in Ref.~\cite{KNT18}. In this strategy, small groups of 
data points consecutive in $s$ are assigned to clusters, after which 
each cluster is assigned an $s$ value according to Eq.~(\ref{clusters}).
The algorithm described in Sec.~\ref{comb} is then applied. The choice of 
clusters can then be varied to find the combination which has both a 
$\chi^2/{\rm dof}$ close to unity and small errors on the sum-rule 
integrals $I^{(w)}_{\rm ex}(s_0)$. A choice of too few data points per
cluster could lead to an erratic point-to-point behavior that would 
not reflect any gain of information, while a choice of too many data 
points per cluster could lead to a loss of information. However, we 
should keep in mind that we are concerned with only two datasets for 
the $2\pi + 4\pi$ contribution to the spectral function that will be 
combined, with one data set (ALEPH) being more precise than the other 
(OPAL). It is then reasonable to consider a combination largely based 
on the ALEPH energy bins, such that the majority of clusters contains 
at least one ALEPH data point. The cluster sizes near the $\rho$ peak 
will be narrower and widen with increasing $s$, since the ALEPH bin 
widths increase with $s$ in the region above the $\rho$ peak.

In order to construct the full covariance matrix $C$ to be used in the 
fit of Eq.~(\ref{chi2}) we first combine the covariances for the $2\p$ and 
$4\p$ $dB/ds$ distributions from ALEPH and OPAL, assuming at this stage 
no correlations between the two data sets. Then BF errors, as well as 
their correlations, are added into the full covariance matrix. This 
introduces correlations between ALEPH and OPAL. With the full covariance 
matrix and a choice of clusters in hand, the fit to Eq.~(\ref{chi2}) can be 
carried out.

%%%%%%%%%%%%%%%%%%%
\begin{figure}[t]
%\vspace*{4ex}
\begin{center}
\includegraphics*[width=10cm]{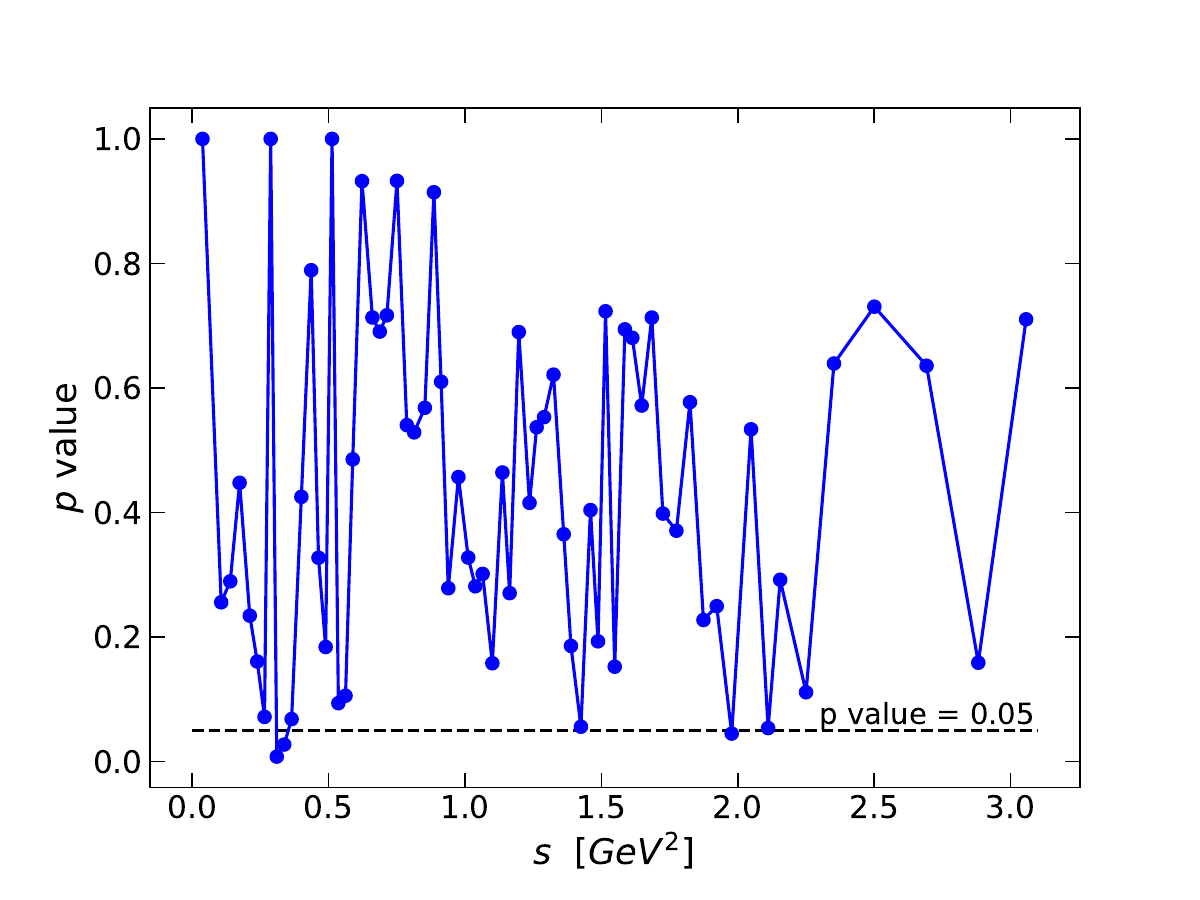}
\end{center}
\begin{quotation}
\floatcaption{figcl2}{{\it $p$-value distribution by cluster for the 
68-cluster combination shown in Fig.~\ref{figcl1}.}}
\end{quotation}
\vspace*{-4ex}
\end{figure}
%%%%%%%%%%%%%%%%%%%

Our final combination contains 68 clusters (a number not too far below 
the 79 bins of the 2013 ALEPH data set) and yields a $\chi^2$ per degree 
of freedom close to the unity:
\begin{equation}
\label{chi2mincl}
\chi^2_{\rm min}/{\rm dof} = 1.144\ ,
\end{equation}
with a good $p$ value of $15\%$. The result of this fit, together with 
the ALEPH and OPAL data sets is shown in Fig.~\ref{figcl1}, where the 
error bars are the non-inflated errors obtained from Eq.~(\ref{clcov}),
and the blue band represents the inflated errors obtained through 
local-$\chi^2$ inflation.\footnote{By ``local-$\chi^2$ inflation'' 
we mean rescaling the errors on those clustered data points with local 
$\chi^2/$dof greater than $1$ so the modified local $\chi^2$/dof values 
become equal to $1$.} 
The local $p$ value for each cluster is shown in Fig.~\ref{figcl2}.    
In the rest of this paper, we will choose to work with non-inflated 
errors. One expects fluctuations in $p$ value when combining these 
two data sets, and the local $p$ value is never unacceptably 
small.\footnote{Only one $p$ value, equal to $0.0077$, is smaller than 1\%.}
There is thus no reason to assume that values of the local $\c^2/$dof signal 
discrepancies between the ALEPH and OPAL data. As we will see in 
Sec.~\ref{strong coupling}, inflating the errors has no effect on $\a_s$.

%%%%%%%%%%%%%%%%%%%
\begin{figure}[t]
%\vspace*{4ex}
\begin{center}
\includegraphics*[width=7.6cm]{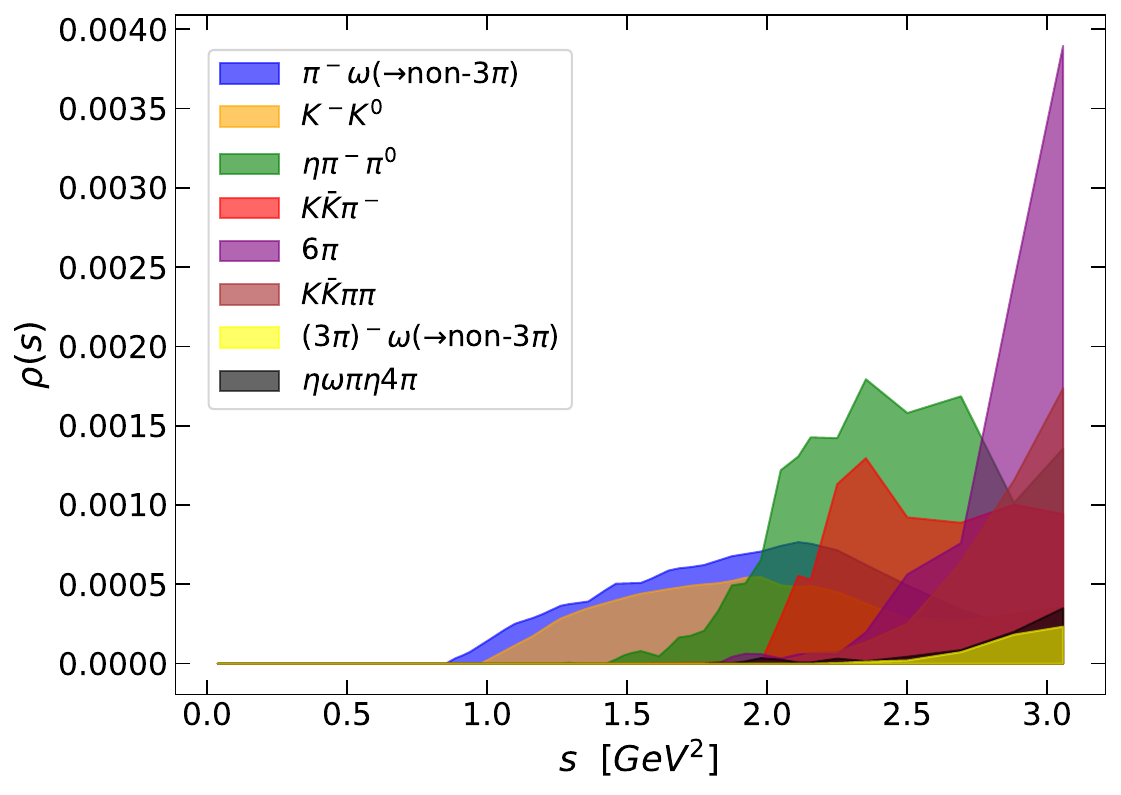}
%\hspace{0.5cm}
\includegraphics*[width=7.4cm]{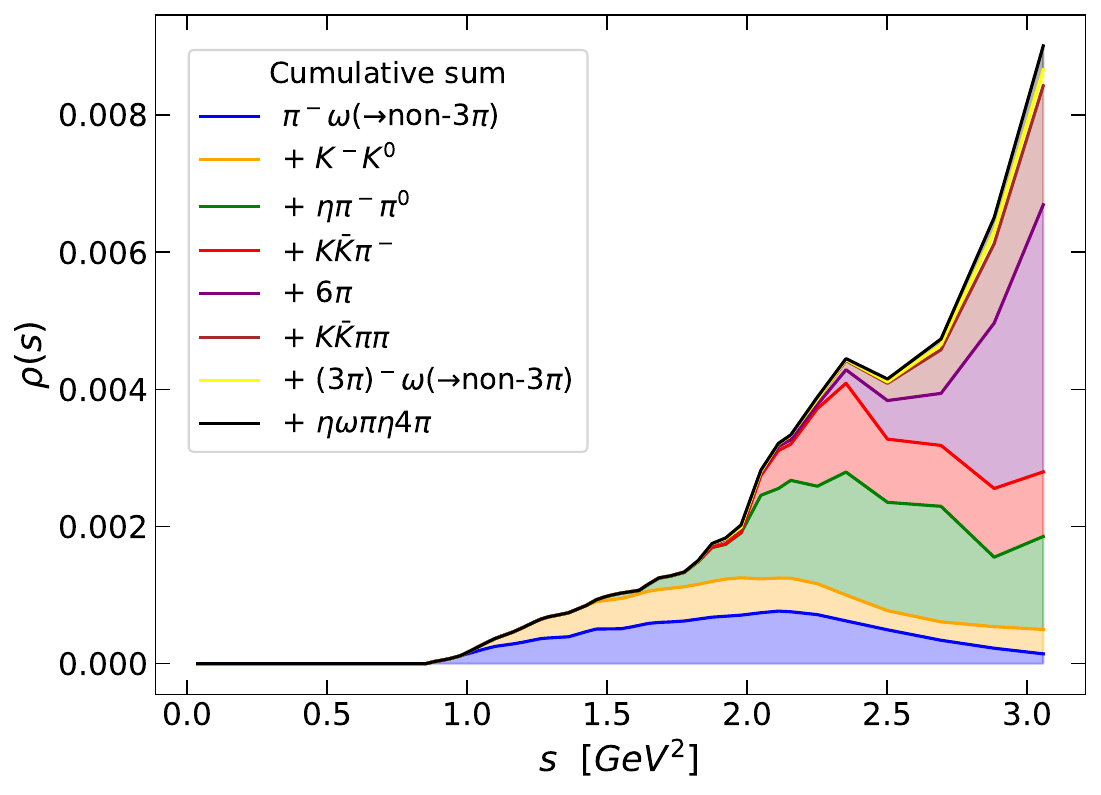}
\end{center}
\begin{quotation}
\floatcaption{figcl3}{{\it Residual modes contributions to the spectral 
function. Left panel: individual modes; right panel: cumulative residual-mode 
sums.}}
\end{quotation}
\vspace*{-4ex}
\end{figure}
%%%%%%%%%%%%%%%%%%%

We explored different choices of the clusters, with the number of 
clusters $N_{\rm cl}$ ranging from 52 to 79 (the latter equaling the 
number of bins in the ALEPH data set). We found no significantly better 
fits with $N_{\rm cl}>68$. This is not surprising, because 68 is not much 
less than the total number of ALEPH data points, which form the more 
precise data set. In fact, we found equally good fits with $N_{\rm cl}<68$.
However, we wish to have sufficiently many values of $I^{(w)}_{\rm exp}(s_0)$ 
available to probe stability of the sum-rule fits to $I^{(w)}_{\rm exp}(s_0)$ 
(\seef\ Sec.~\ref{strong coupling}), and thus do not want to choose 
$N_{\rm cl}$ too small.

With the combined $2\p+4\p$ spectral function in hand, the residual modes 
we need to add to obtain the inclusive spectral function, in order of 
decreasing BF size, are $\pi^- \omega (\to {\rm \mbox{non-}}3\pi)$, 
$K^-K^0$, $\eta \pi^- \pi^0$, $K\bar{K}\pi^-$, $6\pi$, $K\bar{K}2\pi$, 
$(3\pi)^- \omega (\to {\rm \mbox{non-}}3\pi)$ and 
$\eta \omega \pi \eta 4\pi$. In order to add these modes, a linear 
interpolation to the cluster $s^{(m)}$ values is performed individually 
for each mode. In the left panel of Fig.~\ref{figcl3} the individual 
contribution to the spectral function for each residual mode is shown, 
while the right panel shows the cumulative effect, beginning with the 
$\pi^- \omega (\to {\rm \mbox{non-}}3\pi)$ contribution, then adding 
$K^-K^0$, then $\eta \pi^- \pi^0$, and so on. From these figures we also 
see that the residual modes $(3\pi)^- \omega (\to{\rm \mbox{non-}}3\pi)$ 
and $\eta \omega \pi \eta 4\pi$ with the smallest BFs already give 
negligible contributions to the inclusive spectral function total. 
This observation supports the conclusion already noted above that 
omitted contributions from yet-higher-multiplicity modes can be safely 
neglected in the region up to $s=m_\tau^2$ relevant to the current analysis.

Finally, the inclusive spectral function $\rho_{ud;V}(s)$ is given by the 
sum of the contributions from the combined $2\pi + 4\pi$ and interpolated 
residual modes. Figure~\ref{figcl4} shows the individual contributions as 
well as the sum. Notice that the residual modes give only a small 
contribution, which, moreover, is located toward the end of the 
$\tau$-decay spectrum. The final spectral function is displayed 
in Tab.~\ref{tabcl1} with inflated and non-inflated errors for the 
$2\pi + 4\pi$ contribution.\footnote{The associated $68\times 68$
covariance matrix, which we do not display for lack of space, can
be requested from the authors.}

%%%%%%%%%%%%%%%%%%%
\begin{figure}[t]
%\vspace*{4ex}
\begin{center}
\includegraphics*[width=10cm]{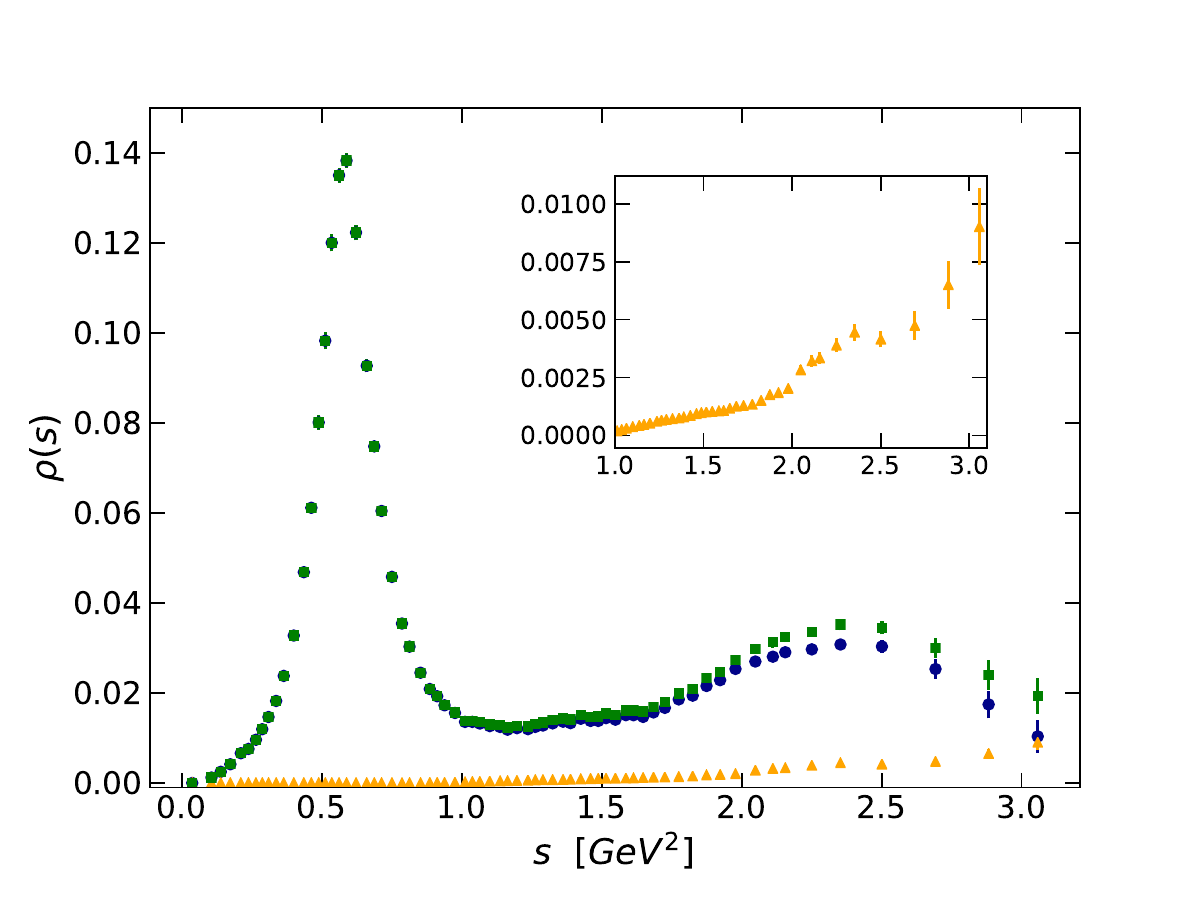}
\end{center}
\vspace*{-6ex}
\begin{quotation}
\floatcaption{figcl4}{{\it The non-strange $V$ combined spectral function.
Total residual-mode contribution (yellow triangles), $2\p+4\p$ 
contribution (blue circles) and the total inclusive spectral 
function (green squares). The inset shows a blow up of the sum of 
residual-mode contributions above $s=1$~{\rm GeV}$^2$. }}
\end{quotation}
\vspace*{-4ex}
\end{figure}
%%%%%%%%%%%%%%%%%%%

\begin{table}[t]
\begin{center}
\begin{tabular}{|c|c||c|c||c|c||c|c|}
\hline
$s$  & $2\pi^2\rho_{ud;V}(s)$ & $s$  & $2\pi^2\rho_{ud;V}(s)$ & $s$  & $2\pi^2\rho_{ud;V}(s)$ & $s$  & $2\pi^2\rho_{ud;V}(s)$ \\
\hline
0.038 &  0.000(00)(00) & 0.106 & 0.024(04)(04) & 0.139 &  0.049(06)(06) & 0.174 & 0.082(07)(07) \\ 
 0.211 & 0.131(09)(07) & 0.238 & 0.150(10)(07) & 0.265 &  0.190(16)(09) & 0.288 & 0.236(11)(11) \\ 
 0.310 & 0.290(26)(10) & 0.337 & 0.360(24)(11) & 0.364 & 0.470(25)(14) & 0.400 & 0.647(16)(16) \\ 
 0.436 & 0.926(20)(20) & 0.463 & 1.208(24)(24) & 0.489 & 1.583(42)(31) & 0.512 & 1.942(37)(37) \\ 
 0.536 & 2.372(63)(38) & 0.562 & 2.668(52)(32) & 0.588 & 2.733(32)(32) & 0.622 & 2.417(33)(33) \\ 
 0.661 & 1.832(29)(29) & 0.688 & 1.478(22)(22) & 0.714 & 1.195(21)(21) & 0.751 & 0.905(19)(19) \\ 
 0.787 & 0.700(17)(17) & 0.814 & 0.599(16)(16) & 0.853 & 0.484(14)(14) & 0.886 & 0.413(12)(12) \\ 
 0.912 & 0.382(10)(10) & 0.939 & 0.343(11)(10) & 0.976 & 0.310(10)(10) & 1.012 & 0.272(10)(10) \\ 
 1.038 & 0.273(10)(10) & 1.065 & 0.267(10)(10) & 1.100 & 0.257(12)(09) & 1.137 & 0.254(09)(09) \\ 
 1.164 & 0.243(10)(09) & 1.197 & 0.251(10)(10) & 1.237 & 0.248(09)(09) & 1.263 & 0.259(09)(09) \\ 
 1.290 & 0.266(10)(10) & 1.325 & 0.276(10)(10) & 1.362 & 0.284(10)(10) & 1.389 & 0.279(13)(10) \\ 
 1.425 & 0.299(18)(11) & 1.460 & 0.291(10)(10) & 1.488 & 0.292(13)(10) & 1.515 & 0.305(11)(11) \\ 
 1.549 & 0.298(15)(11) & 1.586 & 0.319(11)(11) & 1.614 & 0.319(11)(11) & 1.648 & 0.313(11)(11) \\ 
 1.685 & 0.335(12)(12) & 1.726 & 0.355(13)(13) & 1.775 & 0.394(13)(13) & 1.825 & 0.413(15)(15) \\ 
 1.874 & 0.461(17)(14) & 1.923 & 0.488(18)(15) & 1.978 & 0.541(31)(18) & 2.049 & 0.589(20)(20) \\ 
 2.111 & 0.618(46)(24) & 2.156 & 0.640(23)(21) & 2.251 & 0.664(30)(22) & 2.353 & 0.696(25)(25) \\ 
 2.501 & 0.681(29)(29) & 2.692 & 0.594(44)(44) & 2.882 & 0.474(79)(64) & 3.057 & 0.383(78)(78) \\ \hline
\end{tabular}
\floatcaption{tabcl1}{{\it Total inclusive spectral function multiplied 
by $2\p^2$. First errors are inflated errors, while second errors are not 
inflated. We recall that the parton-model value for $2\pi^2\rho_{ud;V}(s)$ 
is $\half$.}}
\end{center}
\end{table}

Having obtained the combined inclusive spectral function, we can compare 
the moments $I^{(w)}_{\rm ex}(s_0)$ which result to those obtained 
using the ALEPH and OPAL versions of $\rho_{ud;V}(s)$. We expect these 
to be consistent with one another, of course, but also expect the errors 
for the combined case to be the smallest. In Table~\ref{tab1} we compare 
values of the $I^{(w)}_{\rm ex}(s_0)$ for $w=w_0$, $w_2$ and $w_3$ at 
$s_0=s_0^*\approx 1.5$~GeV$^2$ and $s_0=s_0^{**}\approx 2.9$~GeV$^2$.
With the binning of the three data sets being somewhat different, the 
central values cannot be directly compared, because the spectral moments 
in the table have been computed at slightly different $s_0$ values on each 
line. However, the errors can be compared, as they vary only very slowly 
with $s_0$. The table thus indicates the gain in precision for spectral 
moments obtained from using the combined spectral function, instead of
the ALEPH or OPAL spectral functions. It should also be borne in mind 
that the errors quoted for the ALEPH and OPAL entries in this table do not 
include additional, difficult-to-quantify systematic uncertainties 
associated with the use by ALEPH and OPAL of MC for the $s$-dependences 
of some of the residual exclusive-mode $\rho_{ud;V}(s)$ contributions. 
This additional systematic is absent from the evaluations of the spectral 
moments using our updated $\rho_{ud;V}(s)$ since the $s$-dependences of the 
numerically relevant contributions from all but the very small residual 
$K\overline{K}\pi\pi$ mode (where, instead, maximally conservative 
experimental constraints are used) are now based on direct experimental 
input.\\

\begin{table}[t]
\begin{center}
\begin{tabular}{|c|l|l|l|}
\hline
&combined & ALEPH & OPAL \\
\hline
$I^{(w_0)}_{\rm ex}(s_0^*)$ & 0.03137(14) & 0.03145(17) & 0.03140(46) \\
$I^{(w_0)}_{\rm ex}(s_0^{**})$ & 0.02952(29) & 0.03133(65) & 0.03030(170) \\
\hline
$I^{(w_2)}_{\rm ex}(s_0^*)$ & 0.02362(10) & 0.02370(13) & 0.02371(23) \\
$I^{(w_2)}_{\rm ex}(s_0^{**})$ & 0.02016(8) & 0.02081(14) & 0.02038(27) \\
\hline
$I^{(w_3)}_{\rm ex}(s_0^*)$ & 0.01774(8) & 0.01783(11) & 0.01788(17) \\
$I^{(w_3)}_{\rm ex}(s_0^{**})$ & 0.01574(6)& 0.01614(8) & 0.01580(14) \\
\hline
\end{tabular}
\end{center}
\floatcaption{tab1}{{\it Comparison of the spectral moments 
$I^{(w)}_{\rm ex}(s_0)$ for $w=w_0$, $w=w_2$ and $w=w_3$ at two values 
of $s_0$, for the combined, the ALEPH, and the OPAL versions of the 
non-strange $V$ spectral function. We choose $s_0^*$ to be the closest 
value larger than or equal to $1.5$~{\rm GeV}$^2$ for each case 
(combined, ALEPH, OPAL), and $s_0^{**}$ to be the closest value
smaller than or equal to $2.9$~{\rm GeV}$^2$ for each case.
Note that, because the values of $s_0^*$ and $s_0^{**}$ are slightly 
different for the three cases, the central values cannot be directly 
compared.  The errors can, however, be compared, as they vary 
slowly with $s_0$.}}
\end{table}

\section{\label{strong coupling} The strong coupling}
In this section, we turn to the determination of $\a_s(m_\t)$ from the 
$V$ non-strange spectral function (and associated covariances) 
obtained in the previous section.  We first briefly outline our strategy 
in Sec.~\ref{strategy}, then present our results in Sec.~\ref{results}. Further
analysis and discussion of these results is contained in Sec.~\ref{analysis}.

\subsection{\label{strategy} Strategy}
The $V$ spectral integrals $I^{(w)}_{\rm exp}(s_0)$ for successive 
values of $s_0$ are highly correlated. Very strong correlations also exist
between $I^{(w)}_{\rm exp}(s_0)$ with different weights $w$. We find that 
it is possible to carry out standard $\chi^2$ fits taking into account all 
correlations when we limit ourselves to the weight $w_0$, while this is not 
the case when fits to combinations of two different weights over the same
interval in $s_0$ are considered.

We will thus carry out two types of fits. First, since we need sensitivity 
to the DV parameters in Eq.~(\ref{ansatz}) \cite{alphas1}, we will always 
include the spectral integrals with the unpinched weight $w_0$ in our fits.
Our most basic fit is a single-weight $\c^2$ fit to 
$I^{(w_0)}_{\rm exp}(s_0)$ over an interval $s_0\in[s_{\rm min},s_{max}]$, 
where we will always choose $s_{max}$ equal to the largest of the cluster 
$s^{(m)}$ values. Since this is a non-linear fit, errors determined from 
the second-derivative matrix at $\c^2_{\rm min}$ are not necessarily very 
meaningful, and we instead determine errors by varying each parameter such 
that $\c^2=\c^2_{\rm min}+1$. In all cases, we find that these errors are 
approximately symmetric, and so take the average of the negative and 
positive errors as our estimate for the error on each parameter. Because 
correlations are fully taken into account, we will also provide the $p$ 
value of these fits, and use this to determine optimal values of $s_{\rm min}$.

In the second class of fit, we carry out combined two-weight fits to 
$I^{(w_0)}_{\rm exp}(s_0)$ and one of the $I^{(w_n)}_{\rm exp}(s_0)$, 
with $n=2$, $3$ or $4$. These fits serve as consistency checks on the 
fits to $I^{(w_0)}_{\rm exp}(s_0)$ alone. For these fits, the combined 
two-weight correlation matrix is too singular to allow a fully
correlated fit to be carried out.  Thus, as in Refs.~\cite{alphas14,alphas1,alphas2}, 
we carry out ``block-diagonal'' fits, using a quadratic form in which all 
correlations between different $s_0$ values for each spectral 
integral are retained, but not those between spectral integrals 
with different weights. All correlations are fully included after 
the fit, when we obtain parameter error estimates for such block-diagonal 
fits by linear error propagation, as summarized in the appendix
of Ref.~\cite{alphas1}. To distinguish it from a true $\c^2$ function, we
will refer to the quadratic form minimized in this type of fit as the 
fit quality $Q^2$. The distinction is relevant since $Q^2$ is, in 
general, not a $\c^2$-distributed quantity. While this makes it more 
difficult to characterize, in a quantitative manner, the quality of 
such a fit, the relative ``goodness'' of two such fits involving the 
same pair of weights, but different values of $s_{\rm min}$, can still 
be assessed by comparing the optimized results for their $Q^2$ 
values per degree of freedom. We emphasize again that the full data 
covariance matrix, including now also correlations between spectral
moments with different weights, is taken into account in the error 
propagation.   

A final complication is caused by the fact that the correlation matrices 
for $I^{(w_3)}_{\rm exp}(s_0)$ and $I^{(w_4)}_{\rm exp}(s_0)$ turn out to 
have very small eigenvalues, for the relevant values of $s_{\rm min}$, 
precluding even the block-diagonal fits with these moments if the full 
set of cluster values $s^{(m)}$ is used for the $s_0$ values in the fit.
It turns out that this problem can be solved by ``thinning'' the data, 
as will be described in more detail in the following subsection. We 
emphasize again that these two-moment fits serve primarily as consistency 
checks on the fully correlated fits to $I^{(w_0)}_{\rm exp}(s_0)$.

\subsection{\label{results} Results}
We begin with fits to $I^{(w_0)}_{\rm exp}(s_0)$, using Eq.~(\ref{cauchy}) with 
the right-hand side of this equation replaced by Eq.~(\ref{sumrule}) with 
$w=w_0=1$. For the reasons explained in Sec.~\ref{FOPTCIPT} we will focus on 
FOPT, although we will briefly quote values obtained using CIPT as well.

%%%%%%%%%%%%%%%%%%%
\begin{figure}[t!]
%\vspace*{4ex}
\begin{center}
\includegraphics*[width=11cm]{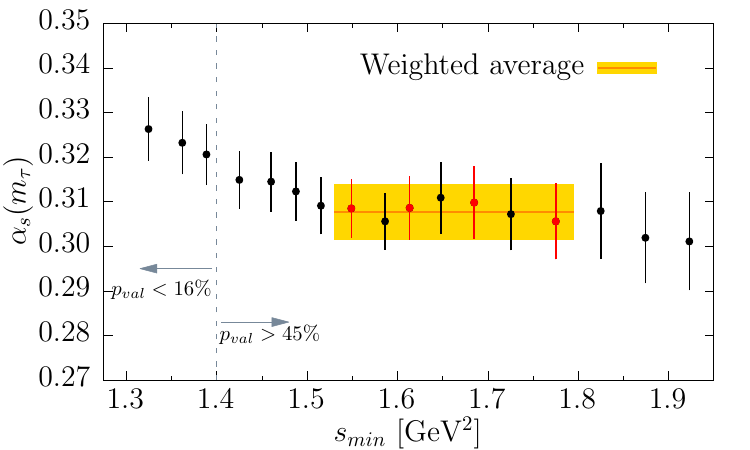}
\end{center}
\vspace*{-4ex}
\begin{quotation}
\floatcaption{fig2}%
{{\it $\a_s(m_\t)$ of Table~\ref{tab2} as a function 
of $s_{\rm min}$. The yellow area correspond to the average reported in 
Eq.~(\ref{parsw0}); this average is computed from the data points indicated 
in red (see text). The thin vertical dashed line separates the regions 
in which the $p$ values shown in Table~\ref{tab2} are smaller than 16\% 
(to the left), from the region where they are larger than 45\%
(to the right).}}
\end{quotation}
\vspace*{-4ex}
\end{figure}
%%%%%%%%%%%%%%%%%%%

%%%%%%%%%%%%%%%%%%%
\begin{figure}[t]
%\vspace*{4ex}
\begin{center}
\includegraphics*[width=8cm]{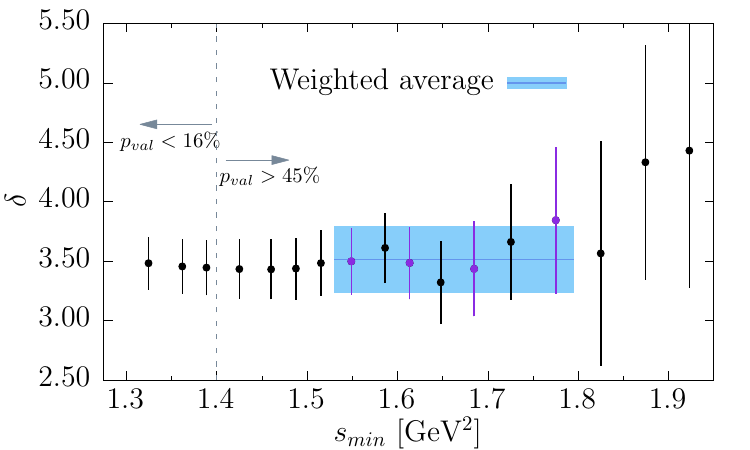}
\hspace{0.2cm}
\includegraphics*[width=8cm]{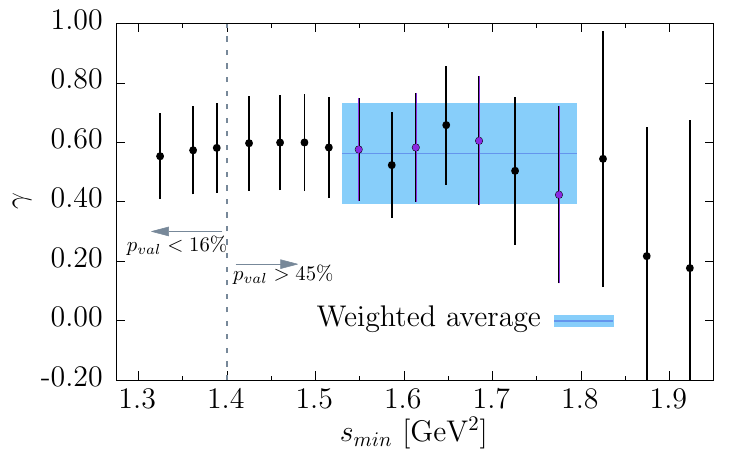}
\vspace{1cm}
\includegraphics*[width=8cm]{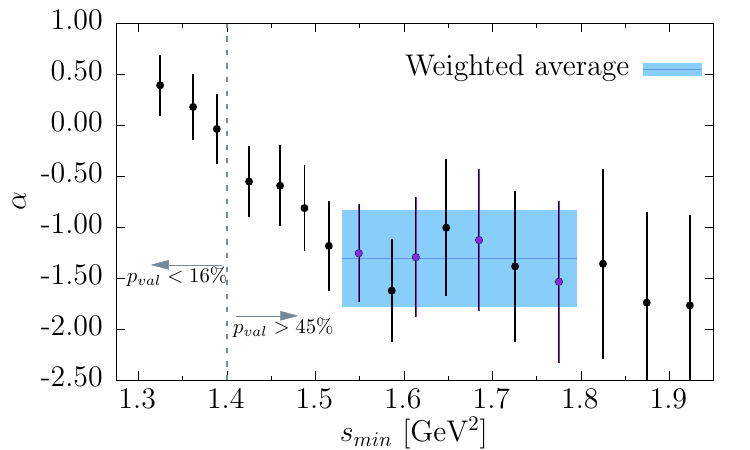}
\hspace{0.2cm}
\includegraphics*[width=8cm]{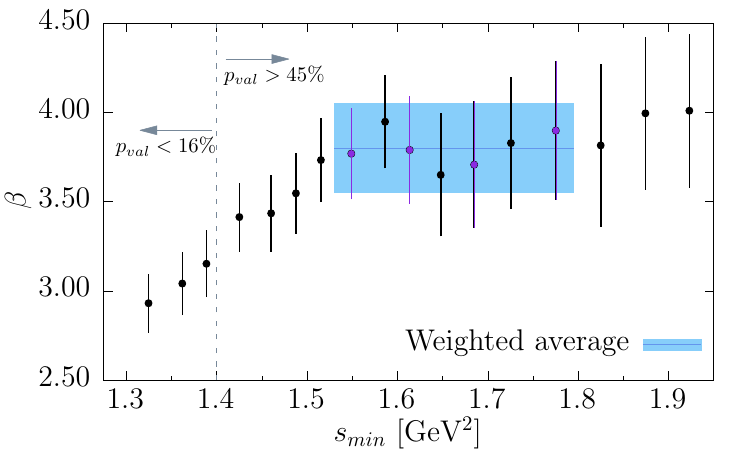}
\end{center}
\vspace*{-8ex}
\begin{quotation}
\floatcaption{fig3}%
{{\it DV parameters of Table~\ref{tab2} as a function 
of $s_{\rm min}$. The blue areas correspond to the averages reported in 
Eq.~(\ref{parsw0}); these averages are computed from the data points indicated 
in purple (see text). $\b$ and $\g$ are in {\rm GeV}$^{-2}$.
The thin vertical dashed line separates the regions in which the $p$ 
values shown in Table~\ref{tab2} are smaller than 16\% (to the left), 
from the region where they are larger than 45\% (to the right).}}
\end{quotation}
\vspace*{-4ex}
\end{figure}
%%%%%%%%%%%%%%%%%%%

\begin{table}[t]
\begin{center}
\begin{tabular}{|l|c|c|l|l|c|c|c|}
\hline
$~~s_{\rm min}$ & $\c^2$/dof & $p$ value &  $~~\a_s(m_\t)$ & $~~~~\,\d$ & $\g$ & $\a$ & $\b$ \\
\hline
1.3246 & 30.71/22 & 0.10 & 0.3263(72) & 3.48(22) & 0.55(14) & \phantom{$-$}0.39(30) & 2.93(17) \\
1.3619 & 28.56/21 & 0.12 & 0.3232(71) & 3.46(23) & 0.57(15) & \phantom{$-$}0.18(32) & 3.04(18) \\
1.3886 & 26.63/20 & 0.15 & 0.3206(70) & 3.45(23) & 0.58(15) & $-$0.04(34) & 3.15(19) \\
1.4251 & 17.96/19 & 0.53 & 0.3149(65) & 3.43(25) & 0.60(16) & $-$0.55(35) & 3.41(19) \\
1.4602 & 17.91/18 & 0.46 & 0.3145(67) & 3.43(25) & 0.60(16) & $-$0.59(40) & 3.43(22) \\
1.4877 & 16.43/17 & 0.49 & 0.3123(67) & 3.44(26) & 0.60(16) & $-$0.81(42) & 3.55(23) \\
1.5154 & 12.69/16 & 0.70 & 0.3091(64) & 3.48(28) & 0.58(17) & $-$1.18(44) & 3.73(24) \\
1.5490 & 12.57/15 & 0.64 & 0.3085(66) & 3.50(28) & 0.58(17) & $-$1.26(48) & 3.77(26) \\
1.5863 & 9.97/14 & 0.76 & 0.3056(64) & 3.61(30) & 0.52(18) & $-$1.62(51) & 3.95(26) \\
1.6136 & 7.65/13 & 0.87 & 0.3084(72) & 3.49(31) & 0.58(18) & $-$1.29(59) & 3.79(30) \\
1.6479 & 6.52/12 & 0.89 & 0.3109(82) & 3.32(35) & 0.66(20) & $-$1.00(68) & 3.65(35) \\
1.6849 & 6.27/11 & 0.85 & 0.3097(83) & 3.43(40) & 0.61(22) & $-$1.13(71) & 3.71(36) \\
1.7256 & 5.71/10 & 0.84 & 0.3072(83) & 3.66(49) & 0.50(25) & $-$1.38(75) & 3.83(37) \\
1.7752 & 5.51/9 & 0.79 & 0.3056(87) & 3.84(62) & 0.42(30) & $-$1.54(80) & 3.90(39) \\
1.8249 & 5.33/8 & 0.72 & 0.308(11) & 3.56(95) & 0.54(44) & $-$1.36(96) & 3.82(47) \\
1.8744 & 4.15/7 & 0.76 & 0.302(11) & 4.3(1.0) & 0.22(45) & $-$1.74(91) & 4.00(44) \\
1.9230 & 4.13/6 & 0.66 & 0.301(12) & 4.4(1.2) & 0.18(52) & $-$1.77(92) & 4.01(45) \\
1.9779 & 0.676/5 & 0.98 & 0.294(10) & 5.4(1.1) & $-$0.19(45) & $-$1.58(85) & 3.97(41) \\
\hline
\end{tabular}
\end{center}
\floatcaption{tab2}%
{{\it Results of fits to $I^{(w_0)}_{\rm exp}(s_0)$ 
employing the combined spectral function with 68 clusters,  with 
$s_{\rm max}=3.0574$~{\rm GeV}$^2$, $s_{\rm min}$ in {\rm GeV}$^2$ and
$\b$ and $\g$ in {\rm GeV}$^{-2}$.}}
\end{table}

The results ofour FOPT fits to $I^{(w_0)}_{\rm exp}(s_0)$ are shown in 
Table~\ref{tab2} for a range of $s_{\rm min}$ values lying between 
$1.3246$ and $1.9779$~GeV$^2$. The dependence of the fit results on 
$s_{\rm min}$ is also displayed in Fig.~\ref{fig2} for $\a_s(m_\t)$, 
and Fig.~\ref{fig3} for the DV parameters. Our first task is to determine
the range of values of $s_{\rm min}$ from which to obtain our best estimate 
of $\a_s(m_\t)$. This selection is based on several observations. First, 
while all $p$ values are acceptable, those for 
$s_{\rm min}\in[1.4251,1.9779]$~GeV$^2$ are very good. Second, all 
fit parameters, including, in particular $\a$ and $\b$, become stable 
for $s_{\rm min}\ge 1.5490$~GeV$^2$, while the fits become distinctly 
less accurate for $s_{\rm min}\ge 1.8249$~GeV$^2$, as the number of 
data points available to the fit becomes smaller.\footnote{For the fit with 
$s_{\min}=1.9779$~GeV$^2$, though the result for the DV parameter
$\g$ is compatible with being positive within errors, the central value 
is negative, which is not allowed theoretically, as it renders the sum 
rule~(\ref{sumrule}) ill defined. We take this as a sign that, at such 
large $s_{\rm min}$, the limited range of $s_0$ available is not sufficient 
for a reliable fit. We omit these results from Figs.~\ref{fig2} 
and~\ref{fig3}.} Finally, we observe that the $p$ value abruptly 
drops for $s_{\rm min}=1.3886$~GeV$^2$ and becomes systematically 
smaller if $s_{\rm min}$ is lowered below this value, signalling the 
expected breakdown of the theory description for low $s_0$. The 
results of our fits are based on the spectral function without 
error inflation in the combined 2$\pi$ +4$\pi$ channels. Inflating 
the errors only produces even higher $p$ values while leaving the 
fit parameters essentially unchanged. For example, for the fit with 
$s_{\rm min} = 1.5863$~GeV$^2$ with error inflation we find a $p$ 
value of 88\% and $\alpha_s(m_\tau)=0.3053(66)$, to be compared 
with 76\% and $\alpha_s(m_\tau)=0.3056(64)$ given in Table~\ref{tab2}. 

Results for the parameters obtained from fits with nearby $s_{\rm min}$ are, 
of course, highly correlated.\footnote{We have calculated the 
correlations between the parameter values obtained from different 
fits with the method discussed in App. A of Ref.~\cite{alphas1}.} 
We will take the correlated average of the parameter values at 
$s_{\rm min}=1.5490$~GeV$^2$, $1.6136$~GeV$^2$, $1.6849$~GeV$^2$ and
$1.7752$~GeV$^2$ as our best estimate for the value of each parameter, 
thinning out the seven points on the interval between $1.5490$~GeV$^2$ 
and $1.7752$~GeV$^2$ to lessen the impact of the very strong correlations 
between the parameter values at neighboring $s_{\rm min}$. With this 
strategy, we find the parameter values (statistical errors only)
\begin{eqnarray}
\label{parsw0}
\a_s(m_\t)&=&0.3077(65)\ ,\qquad(w_0\ ,\ \mbox{FOPT})\\
\d&=&3.51(28)\ ,\nonumber\\
\g&=&0.57(17)\ \mbox{GeV}^{-2}\ ,\nonumber\\
\a&=&-1.31(48)\ ,\nonumber\\
\b&=&3.81(26)\ \mbox{GeV}^{-2}\ .\nonumber
\end{eqnarray}
The points used in obtaining these averages are those marked in red 
and purple in the shaded regions of Figs.~\ref{fig2} and \ref{fig3}, 
respectively.

We note that taking a straight average of the seven values inside the 
yellow window in Fig.~\ref{fig2} between $s_{\rm min}=1.5490$ and 
$1.7752$~GeV$^2$, and taking the smallest parameter error on this 
interval as the error on this average yields $\a_s(m_\t)=0.3080(64)$, 
a result almost identical to that in Eq.~(\ref{parsw0}).\footnote{If 
instead we take a correlated average of five $\a_s$ values taking every 
other point starting from $s_{\rm min}=1.5154$~GeV$^2$, we find a 
value $\a_s(m_\t)=0.3060(62)$. Thus, if we enlarge the window in 
Fig.~\ref{fig2} by one point on each end of the yellow ``plateau,'' we 
find that our fits are stable: the central value moves by about 
one-fourth of the error in Eq.~(\ref{parsw0}), and the error is essentially 
unchanged.   The $p$ value for our correlated averages of values inside these windows (enlarged
or not) are always good.}  We will take the result shown in Eq.~(\ref{parsw0}) as 
our central value, with the slightly larger error shown there.

The fit to $I^{(w_0)}_{\rm exp}(s_0)$ for $s_{\rm min}=1.5490$~GeV$^2$ is 
displayed in the left panel of Fig.~\ref{fig4}. The right panel of the same 
figure shows a comparison of the representation for $\rho_{ud;V}(s)$ obtained 
using the parameters of this fit with the combined experimental result 
obtained in Sec.~\ref{data}.

%%%%%%%%%%%%%%%%%%%
\begin{figure}[t]
%\vspace*{4ex}
\begin{center}
\includegraphics*[width=8cm]{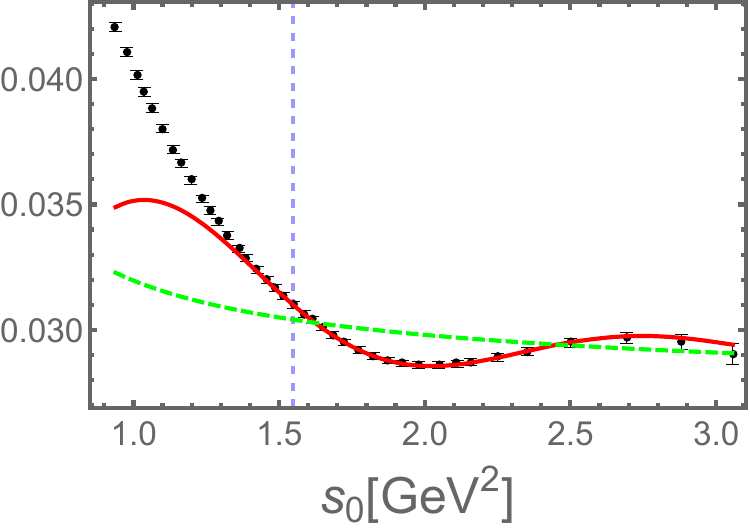}
\hspace{0.3cm}
\includegraphics*[width=7.6cm]{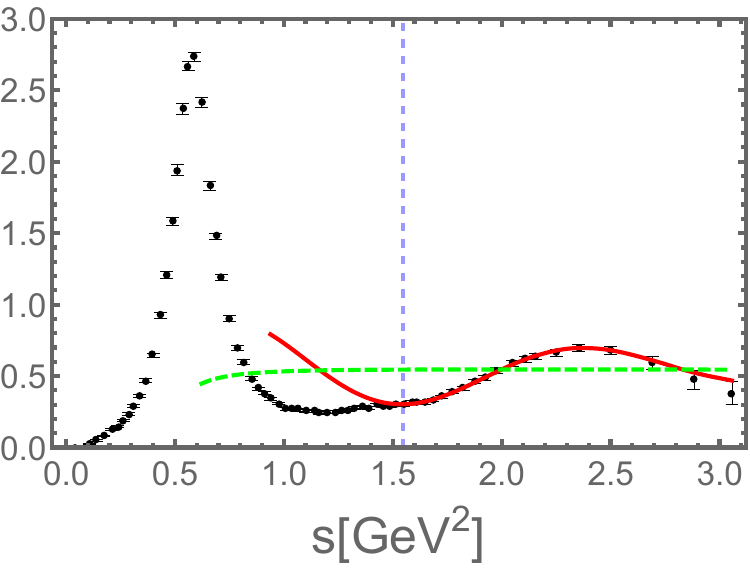}
\end{center}
\begin{quotation}
\floatcaption{fig4}%
{{\it The spectral moment $I^{(w_0)}_{\rm exp}(s_0)$ 
from the fit with $s_{\rm min}=1.5490$~{\rm GeV}$^2$ in Table~\ref{tab2} 
(left panel), and the resulting spectral function (right panel), multiplied 
by $2\p^2$. The black symbols denote data points, the red solid curve the 
fit, and the green dashed curve the OPE part of the fit.}}
\end{quotation}
\vspace*{-4ex}
\end{figure}
%%%%%%%%%%%%%%%%%%%

We have also carried out these fits using the CIPT prescription for the 
$D=0$ perturbative contributions, finding the values (again, statistical 
errors only) 
\begin{eqnarray}
\label{parsw0CI}
\a_s(m_\t)&=&0.3239(87)\ ,\qquad(w_0\ ,\ \mbox{CIPT})\\
\d&=&3.35(28)\ ,\nonumber\\
\g&=&0.65(18)\ \mbox{GeV}^{-2}\ ,\nonumber\\
\a&=&-1.33(49)\ ,\nonumber\\
\b&=&3.80(26)\ \mbox{GeV}^{-2}\ .\nonumber
\end{eqnarray}
We do not show figures equivalent to Figs.~\ref{fig2}, \ref{fig3} and 
\ref{fig4}, as they would look very similar. As is the case in all 
$\t$-based $\a_s$ determinations, the CIPT value is about 5 percent 
larger than the FOPT value. The statistical error on the FOPT  and CIPT 
values are about 2.1 and 2.6 percent, respectively. We note that the 
DV-parameter values are not significantly different between the FOPT and 
CIPT fits.

\begin{table}[t]
\begin{center}
\begin{tabular}{|l|c|l|l|c|c|c|c|}
\hline
$~~s_{\rm min}$ & $Q^2$/dof &  $~~\a_s(m_\t)$ & $~~~~\,\d$ & $\g$ & $\a$ & $\b$ & $10^2c_6$\\
\hline
1.5490 & 26.2/34 &  0.3085(67) & 3.49(28) & 0.58(17) & $-$1.44(52) & 3.85(27) & $-$0.60(12) \\
1.5863 & 22.7/32 &  0.3073(69) & 3.50(29) & 0.58(18) & $-$1.57(58) & 3.92(30) & $-$0.62(13) \\
1.6136 & 18.5/30 &  0.3101(80) & 3.36(31) & 0.65(18) & $-$1.24(68) & 3.76(35) & $-$0.55(17) \\
1.6479 & 15.5/28 &  0.3117(89) & 3.31(35) & 0.67(20) & $-$1.08(78) & 3.68(40) & $-$0.52(20) \\
1.6849 & 15.1/26 &  0.3106(90) & 3.42(40) & 0.62(21) & $-$1.20(81) & 3.74(41) & $-$0.55(20) \\
1.7256 & 13.7/24 &  0.3082(90) & 3.70(48) & 0.49(24) & $-$1.44(85) & 3.85(42) & $-$0.62(19) \\
1.7752 & 13.5/22 &  0.3076(97) & 3.76(61) & 0.46(29) & $-$1.50(91) & 3.88(45) & $-$0.64(21) \\
\hline
\end{tabular}
\end{center}
\floatcaption{tab3}%
{{\it Results of block-diagonal fits to 
$I^{(w_0)}_{\rm exp}(s_0)$ and $I^{(w_2)}_{\rm exp}(s_0)$ employing the 
combined spectral function with 68 clusters,  with 
$s_{\rm max}=3.0574$~{\rm GeV}$^2$,
$s_{\rm min}$ in {\rm GeV}$^2$, $\b$ and $\g$ in {\rm GeV}$^{-2}$, 
$c_6$ in {\rm GeV}$^6$.  
Errors have been computed using linear error propagation.}}
\end{table}

In Table~\ref{tab3}, we present the results for the block-diagonal 
simultaneous fits to $I^{(w_0)}_{\rm exp}(s_0)$ and 
$I^{(w_2)}_{\rm exp}(s_0)$, restricting our attention to the seven 
values of $s_{\rm min}$ used in obtaining the results in Eqs.~(\ref{parsw0}) 
above. Taking a straight average of the seven values between 
$s_{\rm min}=1.5490$~GeV$^2$ and $1.7752$~GeV$^2$, and taking the 
smallest parameter error on this interval as the error on this average,
we find the parameter values (statistical errors only)\footnote{As we 
have seen in the case of $I^{(w_0)}_{\rm exp}(s_0)$, this simplified 
averaging procedure produces a very good approximation to the fully 
correlated average we computed in that case.}
\begin{eqnarray}
\label{parsw2}
\a_s(m_\t)&=&0.3091(69)\ ,\qquad(w_0\ \&\ w_2\ ,\ \mbox{FOPT})\\
\d&=&3.51(29)\ ,\nonumber\\
\g&=&0.58(18)\ \mbox{GeV}^{-2}\ ,\nonumber\\
\a&=&-1.35(58)\ ,\nonumber\\
\b&=&3.81(30)\ \mbox{GeV}^{-2}\ ,\nonumber\\
c_6&=&-0.0059(13) \ \mbox{GeV}^6\ .\nonumber
\end{eqnarray}
These parameter values are in excellent agreement with those in 
Eq.~(\ref{parsw0}); of course, $c_6$ is new.

\begin{table}[t]
\begin{center}
\begin{tabular}{|l|c|l|l|c|c|c|c|c|}
\hline
$~~s_{\rm min}$ & $Q^2$/dof &  $~~\a_s(m_\t)$ & $~~~~\,\d$ & $\g$ & $\a$ & $\b$ & $10^2c_6$ & $10^2c_8$\\
\hline
1.5490 & 3.23/7 &  0.3070(70) & 3.37(34) & 0.66(21) & $-$1.80(62) & 4.05(33) & $-$0.71(12) & 1.23(20) \\
1.5863 & 2.11/7 &  0.3068(74) & 3.37(34) & 0.66(21) & $-$1.88(73) & 4.10(38) & $-$0.72(13) & 1.25(22) \\
1.6136 & 2.19/5 &  0.3097(83) & 3.38(35) & 0.64(21) & $-$1.55(83) & 3.92(43) & $-$0.66(15) & 1.15(28) \\
1.6479 & 1.96/5 & 0.3076(82) & 3.44(41) & 0.63(23) & $-$1.72(83) & 4.01(43) & $-$0.71(15) & 1.24(27) \\
1.6849 & 1.33/5 & 0.3048(80) & 3.62(46) & 0.54(25) & $-$2.13(89) & 4.22(45) & $-$0.78(14) & 1.37(26) \\
1.7256 & 2.03/3 & 0.311(11) & 3.41(70) & 0.63(34) & $-$1.40(1.06) & 3.85(53) & $-$0.65(23) & 1.12(43) \\
1.7752 & 1.81/3 & 0.309(11) & 3.4(1.0) & 0.66(48) & $-$1.52(1.09) & 3.91(54) & $-$0.68(25) & 1.19(50) \\
\hline
\end{tabular}
\end{center}
\floatcaption{tab4}%
{{\it Results of block-diagonal fits to 
$I^{(w_0)}_{\rm exp}(s_0)$ and $I^{(w_3)}_{\rm exp}(s_0)$ employing the
combined spectral function with 68 clusters,  with 
$s_{\rm max}=3.0574$~{\rm GeV}$^2$, $s_{\rm min}$ in {\rm GeV}$^2$, 
$\b$ and $\g$ in {\rm GeV}$^{-2}$, $c_6$ in {\rm GeV}$^6$, $c_8$ in 
{\rm GeV}$^8$. Errors have been computed using linear error propagation.
For each fit, every third value of $I^{(w_0)}_{\rm exp}(s_0)$ and 
$I^{(w_3)}_{\rm exp}(s_0)$ was used, starting from $s_0=s_{\rm min}$. }}
\end{table}

\begin{table}[t]
\begin{center}
\begin{tabular}{|l|c|l|l|c|c|c|c|c|}
\hline
$~~s_{\rm min}$ & $Q^2$/dof &  $~~\a_s(m_\t)$ & $~~~~\,\d$ & $\g$ & $\a$ & $\b$ & $10^2c_6$ & $10^2c_{10}$\\
\hline
1.5490 & 2.89/7 &  0.3069(70) & 3.37(34) & 0.66(21) & $-$1.79(62) & 4.04(33) & $-$0.69(12) & 1.56(33) \\
1.5863 & 1.80/7 &  0.3065(74) & 3.37(34) & 0.66(21) & $-$1.87(73) & 4.09(38) & $-$0.70(13) & 1.60(38) \\
1.6136 & 1.90/5 &  0.3097(83) & 3.38(35) & 0.64(21) & $-$1.52(84) & 3.91(44) & $-$0.64(16) & 1.41(48) \\
1.6479 & 1.70/5 &  0.3076(82) & 3.43(41) & 0.63(23) & $-$1.69(83) & 3.99(43) & $-$0.69(16) & 1.57(48) \\
1.6849 & 1.10/5 &  0.3046(80) & 3.61(46) & 0.55(25) & $-$2.11(89) & 4.21(45) & $-$0.76(15) & 1.83(47) \\
1.7256 & 1.75/3 &  0.311(11) & 3.39(70) & 0.64(34) & $-$1.3(1.1) & 3.82(53) & $-$0.62(24) & 1.34(83) \\
1.7752 & 1.56/3 &  0.309(11) & 3.3(1.0) & 0.68(48) & $-$1.5(1.1) & 3.89(55) & $-$0.65(27) & 1.4(1.0) \\
\hline
\end{tabular}
\end{center}
\floatcaption{tab5}%
{{\it Results of block-diagonal fits to 
$I^{(w_0)}_{\rm exp}(s_0)$ and $I^{(w_4)}_{\rm exp}(s_0)$ 
employing the combined spectral function with 68 clusters, with 
$s_{\rm max}=3.0574$~{\rm GeV}$^2$, $s_{\rm min}$ in {\rm GeV}$^2$, 
$\b$ and $\g$ in {\rm GeV}$^{-2}$, $c_6$ in {\rm GeV}$^6$, $c_{10}$ 
in {\rm GeV}$^{10}$. For each fit, every third value of 
$I^{(w_0)}_{\rm exp}(s_0)$ and $I^{(w_3)}_{\rm exp}(s_0)$ was used, 
starting from $s_0=s_{\rm min}$.}}
\end{table}

In the case of simultaneous block-diagonal fits to $I^{(w_0)}_{\rm exp}(s_0)$ 
and $I^{(w_n)}_{\rm exp}(s_0)$ with $n=3,\ 4$, we find that the correlation
 matrices for the spectral moments with the doubly pinched weights $w_{3,4}$ 
have very small eigenvalues, leading to unstable fits with very large $Q^2$
values (equal to about 16 per degree of freedom for 
$s_{\rm min}\sim 1.6$~GeV$^2$). The smallest eigenvalue in each
such case is around $10^{-10}$, orders of magnitude smaller than 
the smallest eigenvalue for the set of $I^{(w_2)}_{\rm exp}(s_0)$ or 
$I^{(w_0)}_{\rm exp}(s_0)$ integrals, which are around $10^{-6}$ and 
$10^{-5}$, respectively. We find that if we ``thin'' the set of
integrals used in the fit, starting at a given $s_{\rm min}$ and including 
only every second, third, \etc, of the available higher $s_0$, the 
$Q^2/$dof drops rapidly to a value below $1$, and the fit stabilizes 
as we increase the degree of thinning.\footnote{The smallest eigenvalue 
of the correlation matrices for $I^{(w_{3,4})}_{\rm exp}(s_0)$ increases 
to about $10^{-6}$ if we thin by a factor 3.} Tables~\ref{tab4} and 
\ref{tab5} show the results of these fits for the cases $n=3$ and $n=4$, 
always thinning by a factor 3. Comparing the results of tables~ \ref{tab2}, 
\ref{tab3}, \ref{tab4} and \ref{tab5}, we find good consistency among all 
these fits. Using the simplified averaging procedure employed above for 
$n=2$, we find the following parameter values (statistical errors only)
\begin{eqnarray}
\label{parsw3w4}
&\hspace{-0.2cm}w_0\ \&\ w_3\qquad &w_0\ \&\ w_4\nonumber\\
\a_s(m_\t)\ \ =&\ 0.3080(70)\qquad & 0.3079(70)\quad (\mbox{FOPT})\ ,  \\
\d\ \ = &\ \!\!\!\!\!\!3.43(34)\qquad & 3.41(34)\ ,\nonumber\\
\g\ \ =&\ \!\!\!\!\!\!0.63(21)\qquad & 0.64(21)\qquad [\mbox{GeV}^{-2}]\ ,\nonumber\\
\a\ \ =&\ \!\!\!\!\!\!\!\!\!\!\!-1.71(62)\qquad & \!\!\!\!\!-1.68(62)\ ,\nonumber\\
\b\ \ =&\ \!\!\!\!\!\!4.01(33)\qquad & 3.99(33)\hspace{0.82cm}[\mbox{GeV}^{-2}]\ ,\nonumber\\
c_6\ \ =&\ \!\!\!\!\!\!\,-0.0070(12)\qquad & \!\!\!\!\!-0.0068(12)\hspace{0.43cm} [\mbox{GeV}^6]\ ,\nonumber\\
c_8\ \ =&\!\!\!\!\!\!\!\!\!\!\!\,0.0122(20)&\mbox{---}\hspace{1.88cm}[\mbox{GeV}^8]\ ,\nonumber\\
c_{10}\ \ =&\hspace{-2.15cm}\mbox{---} &0.0153(33)\quad[\mbox{GeV}^{10}]\ .\nonumber
\end{eqnarray}
where the first, respectively, second, column corresponds to a simultaneous 
fit to $I^{(w_0)}_{\rm exp}(s_0)$ and $I^{(w_3)}_{\rm exp}(s_0)$, 
respectively, $I^{(w_0)}_{\rm exp}(s_0)$ and $I^{(w_4)}_{\rm exp}(s_0)$. 
The results for those parameters also determined in the earlier fits also
show good consistency with the values obtained in those earlier fits, 
reported in Eqs.~(\ref{parsw0}) and~(\ref{parsw2}). In addition, the values 
for the OPE coefficient $c_6$ shown in Tables~\ref{tab4} and \ref{tab5} 
are consistent with those shown in Table~\ref{tab3}. This constitutes 
an additional non-trivial consistency check.

We end this subsection with a comment. For reasons already explained, we
did not construct the axial equivalent of the new inclusive spectral 
function $\r_{ud;V}$ obtained in Sec.~\ref{data}, and thus did not carry 
out simultaneous fits to the $V$ and $A$ spectral functions. This precludes 
us from testing consistency between vector and axial channels, and from 
carrying out tests based on the Weinberg sum rules, as we did in 
Refs.~\cite{alphas14,alphas1,alphas2}. Here we point out that such tests 
were always successful in the separate analyses of the ALEPH and OPAL 
non-strange inclusive spectral functions. We also note that our
most precise results for $\a_s$ were always obtained from purely 
$V$ channel fits.

\subsection{\label{analysis} Analysis}
To finalize our result for $\a_s(m_\t)$, an estimate is required 
for the error resulting from the use of the four- or five-loop-truncated
perturbation theory. This is obtained following the approach outlined
at the end of Sec.~\ref{FOPTCIPT}. We focus on the single-weight
fit to $I^{(w_0)}_{\rm exp}(s_0)$ with $s_{\rm min}=1.5490$~GeV$^2$. 

It turns out that among the various strategies for estimating this 
error discussed in Sec.~\ref{FOPTCIPT}, varying $c_{51}$ by $\pm 50\%$ 
around thecentral value $c_{51}=283$ yields the largest, and thus most 
conservative, estimate of the truncation error.
Symmetrizing the slightly asymmetric result produces an
uncertainty of $\pm 0.0026$ on  $\a_s(m_\t)$. Alternate error 
estimates based on removing order-$\a_s^5$ terms (\ie, setting
$c_{5m}=0$), or removing both order-$\a_s^4$ and order-$\a_s^5$ terms  
(\ie, setting both $c_{4m}=0$ and $c_{5m}=0$) lead to differences 
equal to or smaller than the differences obtained from the $50\%$ 
variation in $c_{51}$ noted above.   

These observations apply to the perturbative representation for 
$I^{(w_0)}_{\rm th}(s_0)$, and do not necessarily apply to spectral 
moments with other weights. Since moments with different 
weights have different perturbative behaviors \cite{BBJ12,BO20},
we will take the difference between the values of $\a_s$ in 
Eqs.~(\ref{parsw0}) and~(\ref{parsw2}) to reflect an independent source of 
perturbative error. We multiply this difference by a factor two
to take into account the fact that one of the two weights entering
the fit leading to Eq.~(\ref{parsw2}), $w_0$, was also used in obtaining the results quoted in
Eqs.~(\ref{parsw0}). This leads to an additional perturbative uncertainty 
of $\pm 0.0028$ on $\a_s$.

Combining the statistical error of Eq.~(\ref{parsw0}) and the two 
perturbative uncertainties discussed above in quadrature, 
we obtain our final result for $\a_s$ at the $\t$ mass scale:
\begin{eqnarray}
\label{asfinal}
\a_s(m_\t)&=&0.3077\pm 0.0065_{\rm stat}\pm 0.0038_{\rm pert} \\
&=& 0.3077\pm 0.0075\qquad\qquad(n_f=3\ ,\ \mbox{FOPT})\ ,\nonumber
\end{eqnarray}
where the subscripts ``stat" and ``pert" refer to the statistical and 
the perturbative error, respectively.

As we explained in Sec.~\ref{theory}, the $\tau$ scale is sufficiently
low that non-perturbative effects are expected to be potentially 
non-negligible. For $I^{(w_0)}_{\rm exp}(s_0)$ non-perturbative
contributions are generated by DVs, corresponding to the second term 
on the right-hand side of Eq.~(\ref{sumrule}). It is interesting to quantify 
these effects. Even though the moment is dominated by perturbation theory,
we find that the non-perturbative part of $I^{(w_0)}_{\rm th}(s_0)$, 
which is the moment most sensitive to non-perturbative effects,
oscillates with an amplitude typically of order 20\% of the 
$\a_s$-dependent part of the perturbative contribution (obtained by 
subtracting the $\alpha_s$-independent parton-model piece)
with varying $s_0$.

The non-perturbative effect is thus small but significant, and 
this is not surprising. The non-perturbative part accounts for the 
oscillation seen in the spectral function in Fig.~\ref{fig4} (red curve), 
which cannot be accounted for by the OPE (green dashed curve). We believe 
that it is unlikely that any variation of the DV {\it ansatz}~(\ref{ansatz}) 
that does an equally good job of fitting the data would lead to 
a variation in $\a_s$ larger than the error we obtained in 
Eq.~(\ref{asfinal}).\footnote{Contrary to claims in the literature, use
of the truncated-OPE strategy (which ignores DVs, as well as certain 
higher dimension OPE contributions) in sum-rule fits to moments of the 
sum of the $V$ and $A$ spectral functions can lead to systematic 
effects of order 10\% in $\a_s(m_\t)$ \cite{critical}.}
It is clear that the data show the existence of non-zero DVs and, while a 
first-principles derivation from QCD does not exist, the main features of 
a DV {\it ansatz} cannot be taken to be arbitrary. As already pointed out 
in Sec.~\ref{theory}, a minimal set of assumptions, based on commonly 
accepted properties of QCD such as, \eg, Regge behavior, leads to the 
parametrization~(\ref{ansatz}) \cite{BCGMP}.

In fact, we have quantitative information on this issue, from the fits 
involving $I^{(w_n)}_{\rm exp}(s_0)$ with $n=2,3,4$, because of the single 
pinch in $w_2$, and the double pinch in $w_{3,4}$, which suppress DVs at 
different rates. Comparing the values of $\a_s(m_\t)$ in Eqs.~(\ref{parsw2}) 
and~(\ref{parsw3w4}) to the value in Eq.~(\ref{parsw0}), we see that the central 
value of $\a_s(m_\t)$ varies by no more than 0.0004, \ie, 0.13\% of the 
central value, to be compared with the 2.3\% relative error in Eq.~(\ref{asfinal}).
Such variations are much smaller than we would expect were the larger
DV contributions to the $w_0$ sum rule to have been incorrectly represented by
the DV \ansatz\ Eq.~(\ref{ansatz}).

Running the result of Eq.~(\ref{asfinal}) to the $Z$-mass scale using the 
standard self-consistent combination of five-loop running \cite{5loop,5loop2}
with four-loop matching \cite{ScSt,CKS} at the charm and bottom thresholds 
($2m_c(m_c)$ and $2m_b(m_b)$, respectively, with $\overline{\rm MS}$ masses 
from the PDG~\cite{pdg2020}) we obtain the corresponding $n_f=5$ result
\begin{equation}
\label{asZ}
\a_s(m_Z)=0.1171\pm 0.0010\qquad(n_f=5\ ,\ \mbox{FOPT})\ .
\end{equation}
With five-loop running and four-loop matching the uncertainty due to the 
running is very small. If we perform the matching at $m_c(m_c)$ and 
$m_b(m_b)$ we find a shift of just 0.00009, which does not contribute 
to the final uncertainty.

To conclude this section, we compare our new value of $\a_s(m_\t)$ given 
in Eq.~(\ref{asfinal}) with those obtained from analyses of the ALEPH 
data \cite{alphas14}, the OPAL data \cite{alphas2}, and from 
$e^+e^-\to hadrons$ below $2$~GeV \cite{alphasEM}, where the latter was 
based on the combined electroproduction spectral data of Ref.~\cite{KNT18}.
These previously obtained values are
\begin{eqnarray}
\label{asprevious}
\a_s(m_\t)&=&0.325\pm 0.018\qquad(\mbox{OPAL\ data}) \ ,\\
\a_s(m_\t)&=&0.296\pm0.010\qquad(\mbox{ALEPH\ data}) \ ,\nonumber\\
\a_s(m_\t)&=&0.298\pm 0.017\qquad(e^+e^-\mbox{\ data})\ .\nonumber
\end{eqnarray}
Previously, we quoted a weighted average of the two $\t$-based values 
in Eq.~(\ref{asprevious}), of the ALEPH-based and OPAL-based results, 
$\a_s(m_\t)=0.303\pm 0.009$, as our best determination from $\t$ decays. 
This value and the values shown in Eq.~(\ref{asprevious})
are in good agreement with our new, more precise value in Eq.~(\ref{asfinal}).

A direct comparison with other recent determinations of $\a_s$ from 
$\t$ decays \cite{ALEPH13,Pich} is problematic because they are all based 
on the truncated OPE strategy, which was shown in 
Refs.~\cite{critical,EManalysis} to be contaminated by uncontrolled systematic 
effects arising mainly from the neglect of unknown higher-order terms in 
the OPE in~Refs.~\cite{ALEPH13,OPAL,CLEO,Pich}. The values of 
Refs.~\cite{ALEPH13,Pich} are also highly correlated, since they are based on 
the same general strategy and the same ALEPH data set. We note that the 
values of Refs.~\cite{ALEPH13,Pich} are significantly larger than ours 
$\alpha_s(m_Z)=0.1199 \pm 0.0015$, from Ref.~\cite{ALEPH13} and 
$\alpha_s(m_Z)=0.1197 \pm 0.0015$, from Ref.~\cite{Pich}.

\section{\label{conclusion} Conclusion}

The determination of the strong coupling from hadronic $\t$ decays has the 
potential to provide one of the most precise values among the many 
determinations from different methods that have appeared in the literature. 
It thus makes sense to aim for a determination from the combined experimental 
information available, and this is what we set out to do in this paper. This 
led us to construct a new non-strange vector, isovector spectral 
function, which is presented in Table~\ref{tabcl1} and Fig.~\ref{figcl4}.

In order to construct this spectral function, we combined the 
$\t\to\p^-\p^0\n_\t$, $\t\to2\p^-\p^+\p^0\n_\t$ and $\t\to\p^-3\p^0\n_\t$ 
experimental data available from the ALEPH and OPAL collaborations, using 
the method employed before in Ref.~\cite{KNT18}. The sum of these contributions 
constitutes 98\% of the spectral function as measured by branching fraction.

Details of the contributions from the remaining exclusive channels,
a number of which were estimated using Monte-Carlo, were not provided
by ALEPH or OPAL. We have replaced the estimates for these residual-mode
contributions using recent $\tau$ results for the $K^- K^0$ mode
and the large amount of data now available, via CVC, from electroproduction 
experiments for the remaining residual modes, with conservative estimates 
of the systematic errors associated with this approach. As measured
by the spectral moments shown in Table~\ref{tab1}, this leads to a more 
accurate determination of the spectral function $\rho_{ud;V}(s)$, 
especially in the upper part of the $\t$ kinematic range. This is a
consequence of the fact that electroproduction data are not 
kinematically limited near the $\t$ mass. We emphasize that the
inclusive spectral function which results is a sum of $s$-dependent 
exclusive-mode contributions, all of which are now obtained from experiment 
and none of which require Monte-Carlo input any more.

One of the most important applications of this new combined data set is a 
determination of the strong coupling $\a_s$ at the $\t$ mass scale. 
We employed previously developed methods using finite-energy sum rules to 
extract a new estimate of the $\overline{\rm MS}$ value of $\a_s(m_\t)$ 
from these data, which, when evolved to the $Z$ mass scale, produces a 
five-flavor result with an estimated precision of about 0.8\%. Our final 
result is $\a_s(m_Z)=0.1171\pm 0.0010$. We also revisited the 
question of how to best estimate the effect of truncating the perturbative 
expansion for the moments involved in these sum rules, arguing that, for a
direct comparison with values obtained from other methods, the fixed-order 
resummation scheme is the appropriate one. Because the perturbative Adler 
function is known to a high order, $\a_s^4$, and the QCD $\b$ function is 
known to an even higher order, $\a_s^5$, we arrived at a systematic error 
reflecting the use of perturbation theory which is quite small, 1.2\% at 
the $\t$ mass. We did not carry out an analysis of the perturbative error 
for CIPT, because of the issues discussed in Sec.~\ref{FOPTCIPT}.
For the same reasons, we emphasize that averaging the FOPT and CIPT values, 
or taking their difference as a measure of the perturbative error, would be 
misleading.

In order to carry out the analysis, we had to rely on the DV 
{\it ansatz}~(\ref{ansatz}), and this introduces a model element into our 
framework. We note in this regard that, while no derivation of Eq.~(\ref{ansatz}) 
from QCD exists, there are strong theoretical arguments supporting this 
{\it ansatz} based on commonly accepted conjectures about the spectrum of 
QCD \cite{BCGMP}, \seef\ Sec.~\ref{FESR}. The consistency of the results of 
the fits presented above, which employ spectral moments with varying 
degrees of DV suppression, moreover, supports the consistency of our 
approach, as discussed in more detail in Sec.~\ref{analysis}.

Our approach can be subjected to further tests when more precise 
$\t$-decay data become available. With the strategy employed above,
more precise data for just the low-multiplicity decay channels 
$\t\to\p^-\p^0\n_\t$, $\t\to2\p^-\p^+\p^0\n_\t$ and $\t\to\p^-3\p^0\n_\t$ 
would produce further improvements to the inclusive non-strange
vector isovector spectral function and allow for much higher precision 
tests of our framework. High-precision data for $\t\to\p^-\p^0\n_\t$ 
are, in fact, already available \cite{BELLE}, implying that similar
data for the two four-pion channels would potentially have a 
significant impact on the determination of $\a_s$ from non-strange 
hadronic $\t$ decays.

Finally, the new spectral function can also be used in other applications.
It can be used to put constraints on low-energy constants in chiral 
perturbation theory, with next-to-next-to-leading-order theoretical 
representations being available \cite{GK1,GK2,ABT}. Additional
low-energy constants would become accessible were an updated non-strange
axial vector spectral function to be obtained. Such an update is, however,
more difficult, given the absence of any analogue of the electroproduction 
data used in improving many of the vector residual-mode contributions. 
We leave the consideration of the axial channel to future work.

\vspace{3ex}
\noindent {\bf Acknowledgments}
\vspace{3ex}

We thank Alex Keshavarzi for discussions. We thank the ICTP South 
American Institute for Fundamental Research (SAIFR) at IFT-UNESP, where 
this work was initiated, for hospitality.
DB is supported by the S\~ao Paulo Research Foundation (FAPESP) 
Grant No.~2015/20689-9 and by CNPq Grant No.~309847/2018-4.
MG and WS are supported by the U.S.\ Department of Energy,
Office of Science, Office of High Energy Physics, under Award No.
DE-SC0013682.
MVR is supported by FAPESP grant No.~2019/16957-9.
KM is supported by a grant from the Natural Sciences and Engineering 
Research Council of Canada.
SP is supported by CICYTFEDER-FPA2017-86989-P and by Grant No. 2017 SGR 1069.
IFAE is partially funded by the CERCA program of the Generalitat de Catalunya.
%%%%%%%%%%%%%%%%%%%%%%%%%%%

\end{document}